\documentclass[tighten,iop,apj,numberedappendix]{emulateapj}

\usepackage{amsmath}
\usepackage{amsfonts}
\usepackage{amssymb}
\usepackage{graphicx}
\usepackage{color}

\usepackage{natbib}

\usepackage{url}

\newcommand{\pd}{\partial}
\newcommand{\ud}{\ensuremath{\mathrm{d}}}

\begin{document}

\title{A New Multi-Dimensional General Relativistic Neutrino
Hydrodynamics Code of Core-Collapse Supernovae \\
III. Gravitational Wave Signals from Supernova Explosion Models}

\author{Bernhard~M\"uller, Hans-Thomas~Janka, Andreas~Marek}

\begin{abstract}
We present a detailed theoretical analysis of the gravitational-wave
(GW) signal of the post-bounce evolution of core-collapse supernovae
(SNe), employing for the first time relativistic, two-dimensional (2D)
explosion models with multi-group, three-flavor neutrino transport
based on the ray-by-ray-plus approximation. The waveforms reflect the
accelerated mass motions associated with the characteristic
evolutionary stages that were also identified in previous works: A
quasi-periodic modulation by prompt postshock convection is followed
by a phase of relative quiescence before growing amplitudes signal
violent hydrodynamical activity due to convection and the standing
accretion shock instability during the accretion period of the stalled
shock. Finally, a high-frequency, low-amplitude variation from
proto-neutron star (PNS) convection below the neutrinosphere appears
superimposed on the low-frequency trend associated with the aspherical
expansion of the SN shock after the onset of the
explosion. Relativistic effects in combination with detailed neutrino
transport are shown to be essential for quantitative predictions of
the GW frequency evolution and energy spectrum, because they determine
the structure of the PNS surface layer and its characteristic g-mode
frequency. Burst-like high-frequency activity phases, correlated with
sudden luminosity increase and spectral hardening of electron
(anti-)neutrino emission for some $10\, \mathrm{ms}$, are discovered
as new features after the onset of the explosion. They correspond to
intermittent episodes of anisotropic accretion by the PNS in the case
of fallback SNe.  We find stronger signals for more massive
progenitors with large accretion rates.  The typical frequencies are
higher for massive PNSs, though the time-integrated spectrum also
strongly depends on the model dynamics.
\end{abstract}

\keywords{supernovae: general---neutrinos---radiative
  transfer---hydrodynamics---gravitation--gravitational
  waves}

\section{Introduction}
\label{sec:intro}
Core-collapse supernovae have been considered as a source of
gravitational waves (GWs) since the 1960s \citep{weber_66}. While they
may have been ousted by compact binary mergers as the most promising
source for the first direct detection of GWs by now, the prospective
GW signal from a nearby supernova may still provide enormously
important clues about the dynamics in the supernova core, shed light
on the nature of the explosion mechanism, and could possibly provide
constraints for the equation of state (EoS) of neutron star matter.
If supernovae are to become a fruitful subject of the future field of
observational GW astronomy, this will require reliable waveform
predictions, and these can only be based on simulations that
accurately capture the evolution from the collapse through several
hundreds of milliseconds of accretion by the stalled shock front
into the explosion phase -- a formidable challenge considering the
host of different factors (neutrino transport, multi-dimensional
hydrodynamical instabilities, general relativity, EoS physics, etc.)
that influence the dynamics.

There is a number of possible scenarios for GW emission from
supernovae. \emph{Rotating} progenitors already produce a signal
during the phases of collapse and bounce due to the breaking of
spherical symmetry, a scenario which has long received interest from
the numerical relativity community in particular (see
\citealp{ott_08_b} for an extensive review). State-of-the art
predictions of the GW signal come from 2D and 3D general relativistic
simulations
\citep{ott_06_a,dimmelmeier_07_a,dimmelmeier_08,abdikamalov_10,ott_12}
employing a parametrized ``deleptonization scheme'' for the
core-collapse phase \citep{liebendoerfer_05_b}, and, most recently, a
neutrino leakage treatment for the post-bounce phase
\citep{ott_12}. The accuracy of this current approach is yet to be
tested against self-consistent models including neutrino transport,
but it is conceivable that the dynamics of rotational collapse can be
captured reasonably well even with a simplified neutrino
treatment. However, the fact that most supernova progenitors are
believed to rotate rather slowly \citep{heger_05} implies that the
signal from rotational core bounce may be weak and difficult to
detect.

The situation is different for the GWs produced by the different
hydrodynamical instabilities that develop during the post-bounce phase
also in non-rotating progenitors, such as convection in the
proto-neutron star (as first pointed out by \citealt{epstein_79}), in
the neutrino-heated hot-bubble region
\citep{bethe_90,herant_92,herant_94,burrows_95,janka_96,mueller_97},
the standing accretion-shock instability (``SASI'',
\citealp{blondin_03,blondin_06,foglizzo_06,ohnishi_06,
  foglizzo_07,scheck_08,iwakami_08,iwakami_09,fernandez_09,fernandez_10}).

Regardless of the actual nature of the explosion mechanism, the
evolution during the post-bounce phase depends crucially on the
effects of neutrino heating and cooling and thus requires an elaborate
treatment of the microphysics and in particular of the neutrino
transport. Because of this constraint, studies addressing the GW
signal from the first several hundreds of milliseconds of the
post-bounce evolution have been limited either to the Newtonian
approximation or the ``pseudo-Newtonian effective potential'' approach
\citep{marek_06} until now. Since the 1990s, several authors have
investigated the problem in 2D and 3D, mostly relying on parametrized
approximations for the neutrino heating and cooling
\citep{mueller_97,kotake_07,murphy_09,kotake_09}, the IDSA method
\citep{scheidegger_10}, or on gray neutrino transport schemes
\citep{fryer_04,mueller_e_12}.  In addition, gravitational waveforms
from state-of-the art multi-group neutrino hydrodynamics simulations
of core-collapse supernovae have become available during the recent
years \citep{mueller_04,ott_07,marek_08,yakunin_10}. These different
(pseudo-)Newtonian studies have by now established the qualitative
features of the GW signal from the post-bounce phase: During the first
several tens of milliseconds, prompt post-shock convection gives rise
to a quasi-periodic signal in the range around $100 \ \mathrm{Hz}$
\citep{marek_08,murphy_09,yakunin_10}, which is followed by a period of
reduced GW activity until hot-bubble convection and the SASI become
vigorous and lead to the emission of a stochastic signal with typical
frequencies of several hundreds of $\mathrm{Hz}$
\citep{mueller_04,marek_08,murphy_09}. After the onset of the
explosion, asymmetric shock expansion may produce a ``tail'', i.e.\ a
growing offset, in the GW amplitude \citep{murphy_09,yakunin_10,
  mueller_e_12}, and a high-frequency signal from proto-neutron star
convection starts to appear. Tail-like signals may also result from
anisotropic neutrino emission
\citep{mueller_04,marek_08,yakunin_10,mueller_e_12} powered by
accretion downflows onto the proto-neutron star.

However, the accuracy of the Newtonian approximation and even of the
``effective potential'' approach, which have formed the basis of GW
predictions for the post-bounce phase so far, is limited. Both 1D
studies \citep{baron_89,bruenn_01,lentz_12} as well as the recent 2D
models of our own group \citep{mueller_12} and
  exploratory GR simulations with simplified transport in 3D
  \citep{kuroda_12} have demonstrated that GR effects have a
non-negligible impact on the neutrino emission, the shock position,
and the heating conditions in the supernova core. It is conceivable
that GR affects the GW signal to a similar degree. To address this
question, we present gravitational waveforms from axisymmetric (2D)
general relativistic simulations (using the xCFC approximation of
\citealt{cordero_09}) of the post-bounce phase including multi-group
neutrino transport for the first time. We compare our results with
predictions from Newtonian and pseudo-Newtonian models. We discuss the
relativistic GW signals of six different progenitors with zero-age
main sequence masses ranging from $8.1 M_\odot$ to $27 M_\odot$ (among
them five explosion models), part of which have already been studied
in \citet{mueller_12} (paper~II) and \citet{mueller_12b}.  For a $15
M_\odot$ star, two simulations using the Newtonian approximation and
the effective potential approach provide the basis for diagnosing GR
effects and for working out the reason of systematic differences where
possible. We also consider a $15 M_\odot$ model with slightly
simplified neutrino rates from \citet{mueller_12}. Although the
  models analyzed in this paper represent the state-of-the-art in 2D
  core-collapse supernova simulations with energy-dependent neutrino
  transport, we emphasize that the quest towards better explosion
  models -- ultimately from 3D GR simulations -- is continuing
  apace. For an overview of the current status, the remaining
  uncertainties of the current generation of models, and future
  directions in the field the reader should refer to
  \citet{mueller_12} as well as recent review articles
  \citep{janka_12,janka_12b,kotake_12,burrows_12b}.

Our paper is structured as follows: In Section~\ref{sec:numerics}, we
briefly summarize the numerical methods and the model setup, which is
laid out in greater detail in \citet[paper~I and
  II]{mueller_10,mueller_12}. In Section~\ref{sec:extraction}, we
describe the methods used for extracting and analyzing the
gravitational wave signals.  The wave signals produced during the
different phases of the post-bounce evolution -- the early phase of
shock propagation and prompt post-shock convection, the steady-state
accretion phase, and the explosion phase -- are discussed in
Section~\ref{sec:general_features}  on the basis of two
exemplary explosion models. The dependence of the gravitational wave
emission on the progenitor model is discussed separately in
Section~\ref{sec:progenitors}. Finally, we summarize the most salient
features of our relativistic waveforms in
Section~\ref{sec:conclusions}.  Important technical issues relied upon
in our paper, such as the derivation of a modified GW quadrupole
formula, analytic expressions for the GW signal produced by an
aspherical shock front, and the Brunt-{V\"ais\"al\"a} (buoyancy)
frequency in GR are treated in Appendices~A, B and C.
Note that different from the rest of the paper we use geometrical
units ($c=G=1$) in the Appendix.

\begin{figure*}
  \plottwo{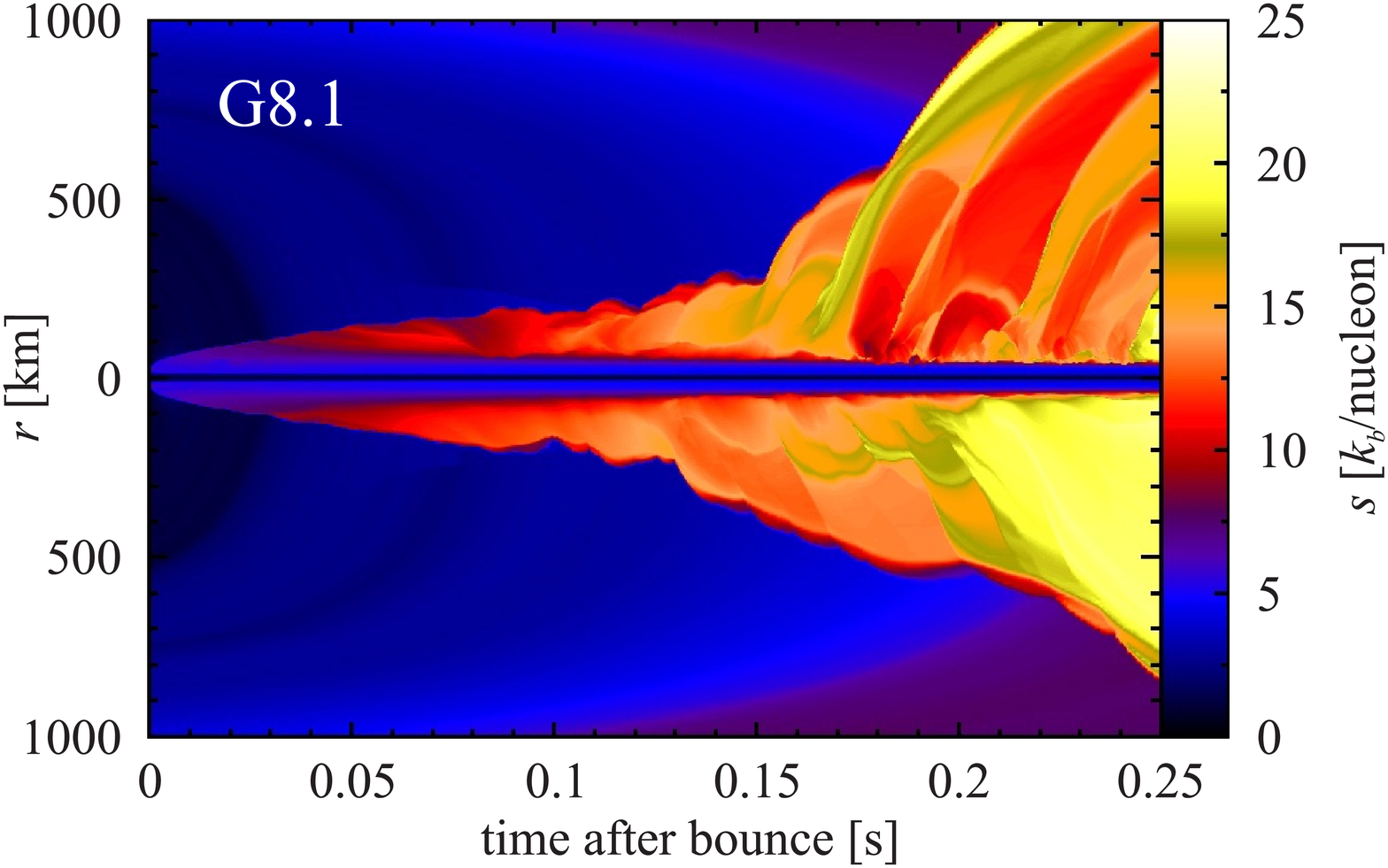}{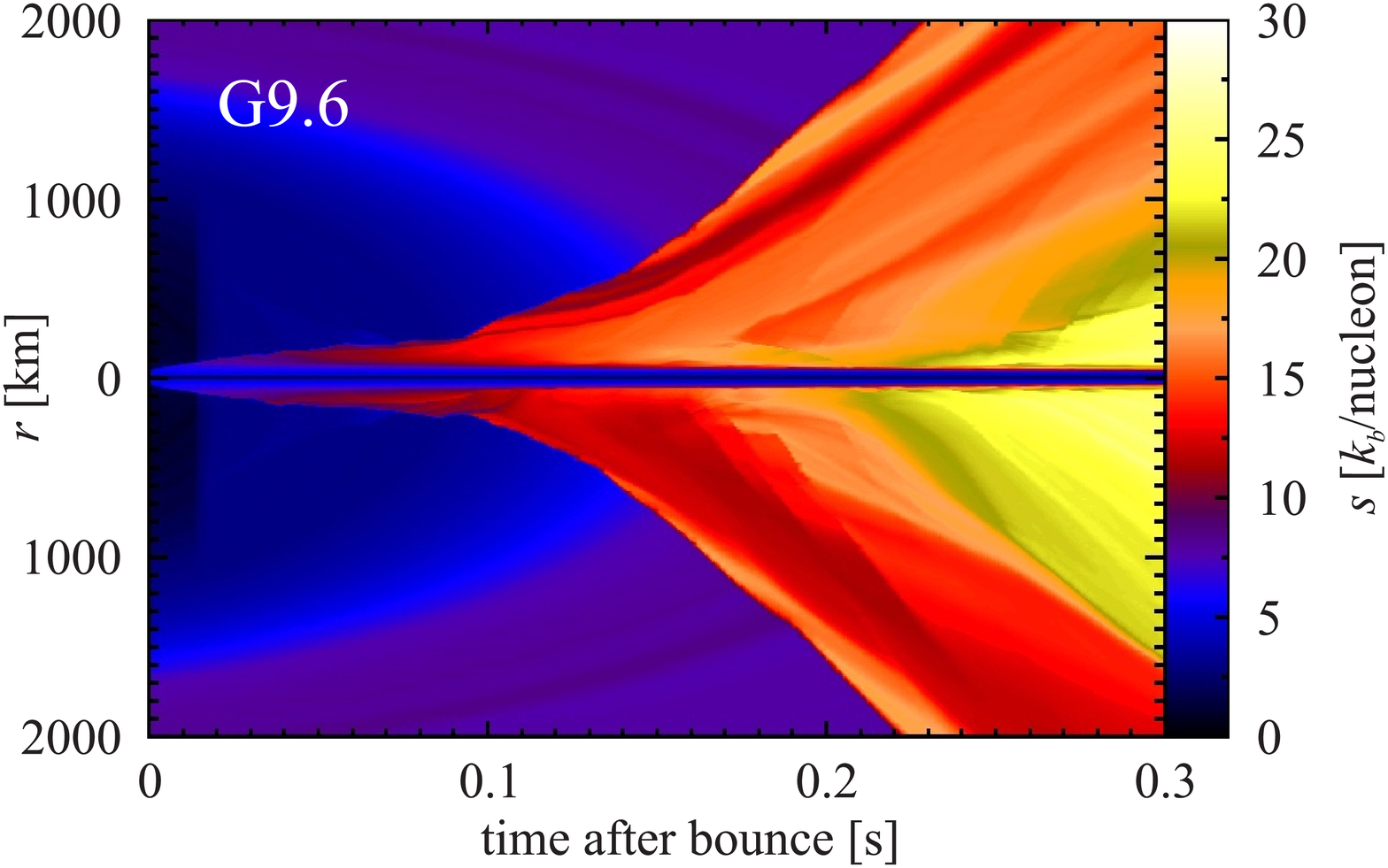}
  \plottwo{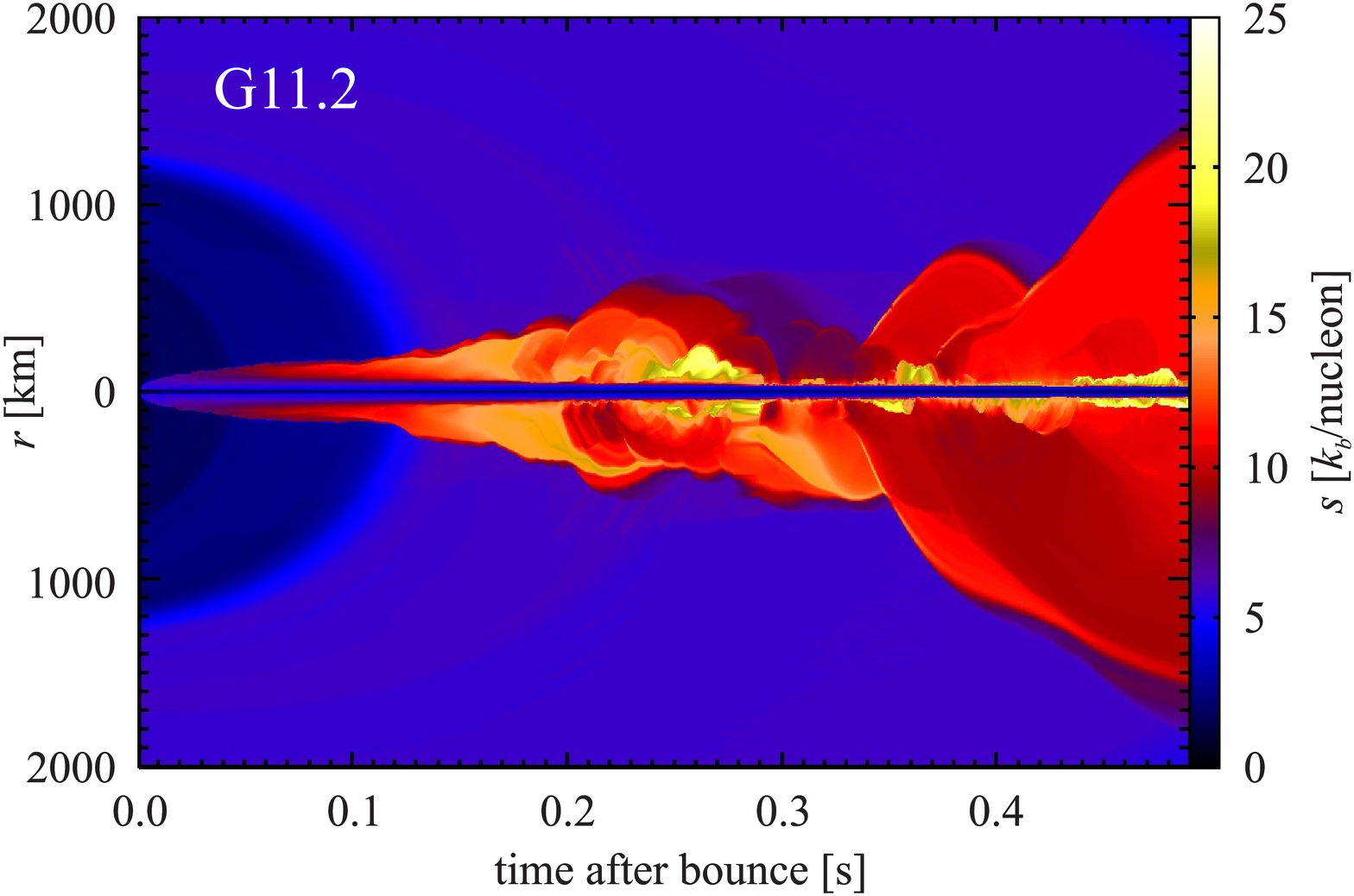}{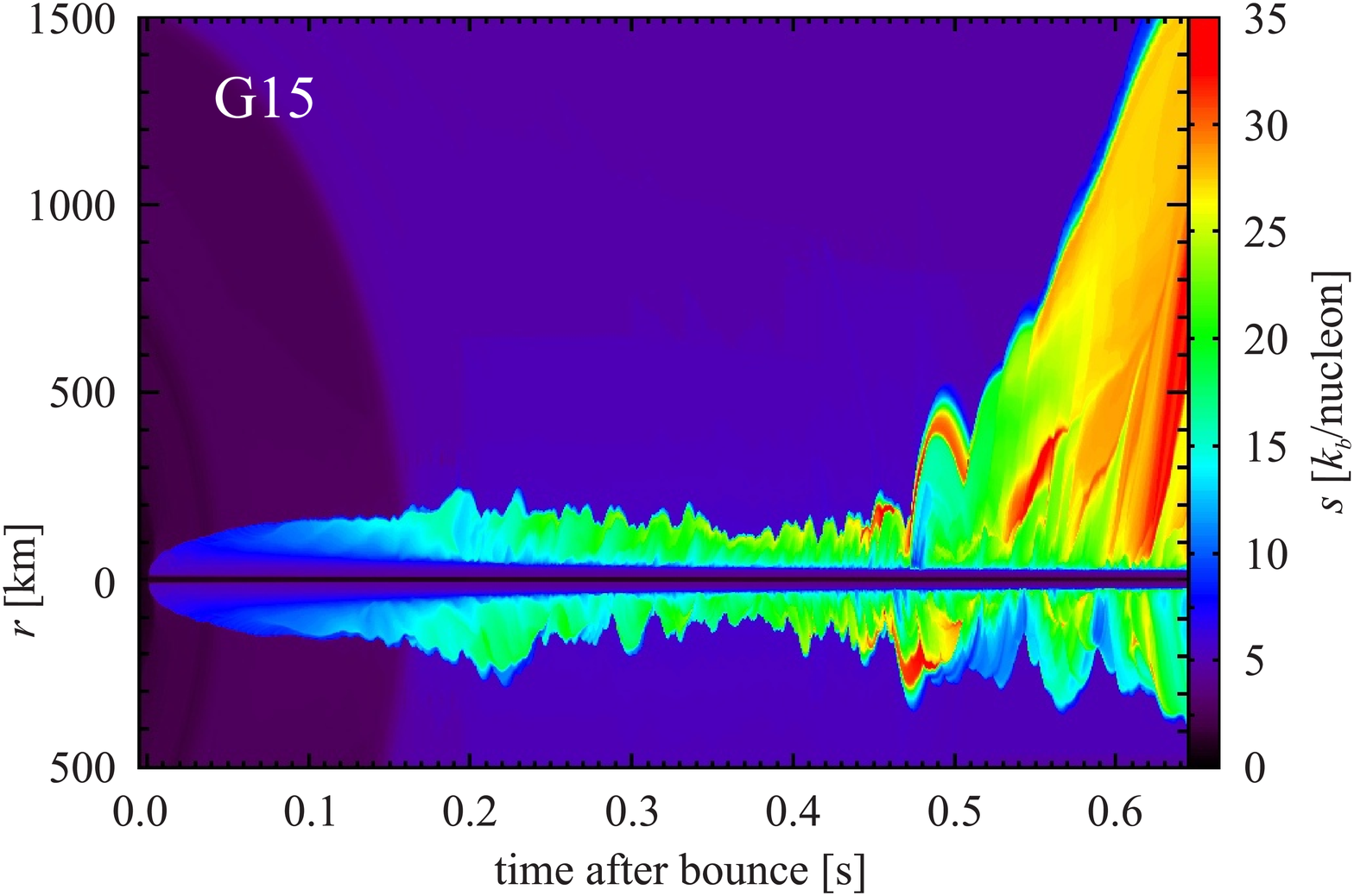}
  \plottwo{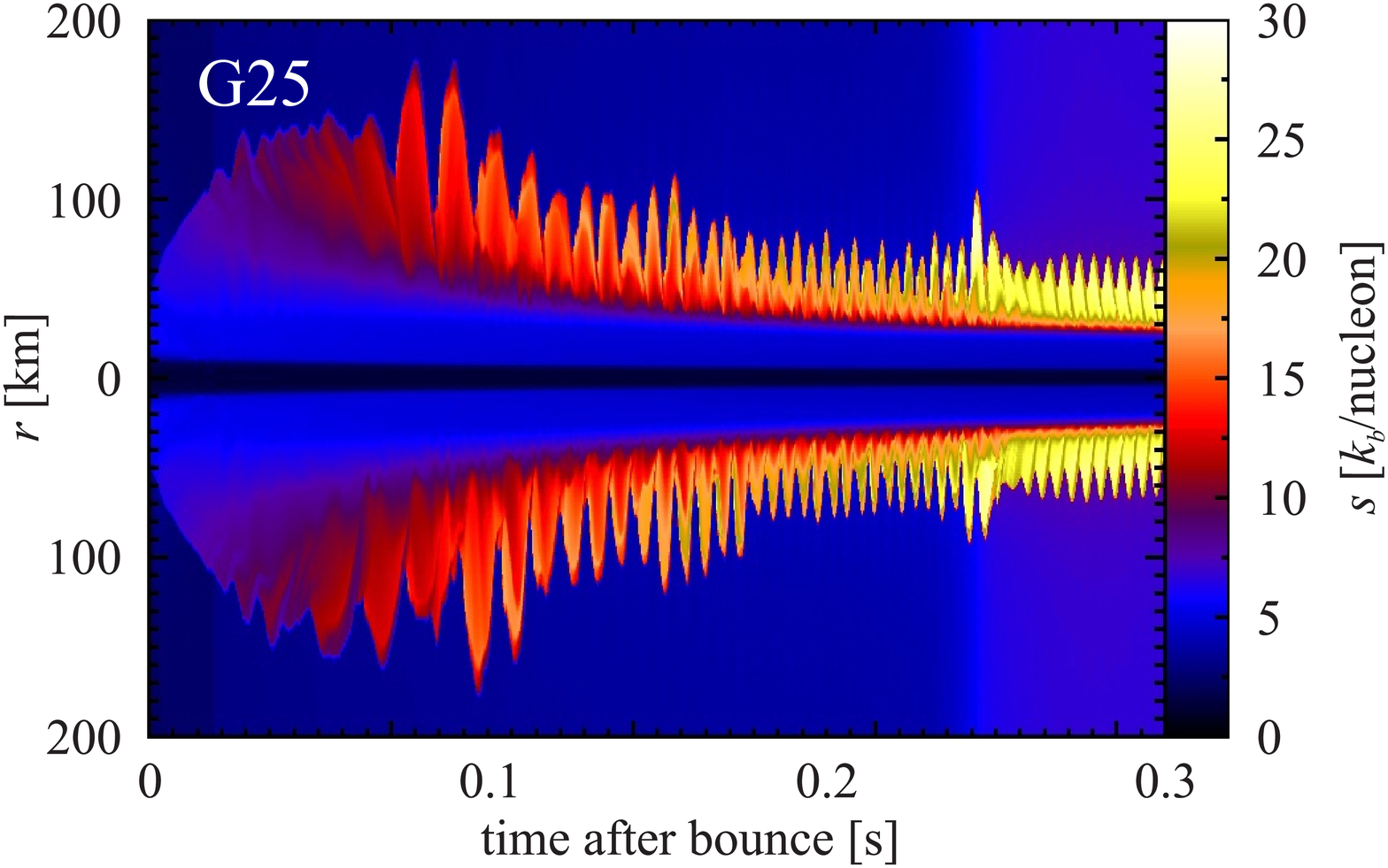}{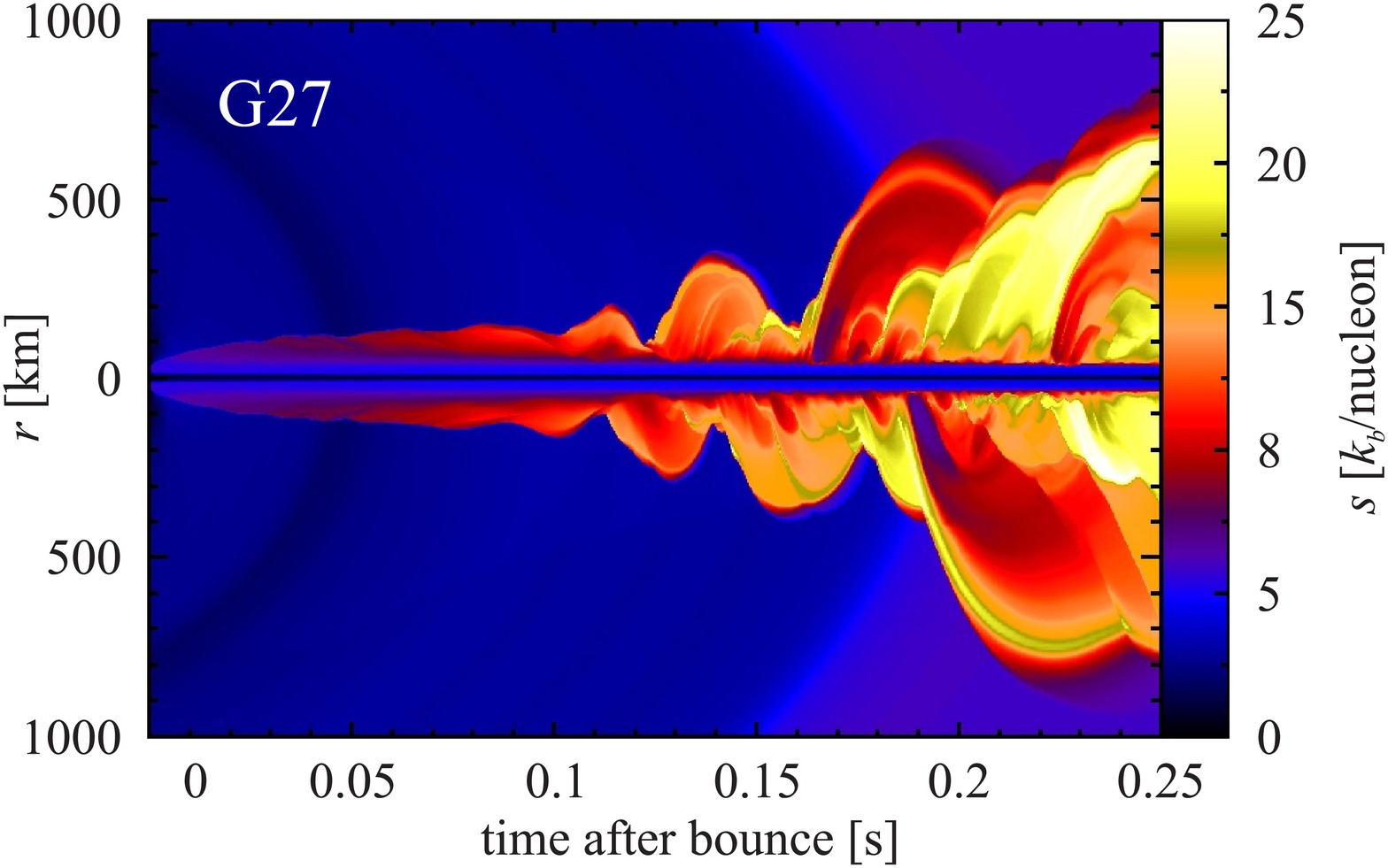}
  \caption{ Entropy along the north and south polar axis as a function
    of time for the relativistic simulations G8.1, G9.6, G11.2, G15,
    G25, and G27 (top left to bottom right). The shock trajectory is visible as
    a discontinuity between the darker violet and blue tones in the pre-shock region
    and lighter colors (red, green, yellow) in the post-shock region.
  \label{fig:model_dynamics}
  }
\end{figure*}

\section{Numerical Methods and Model Setup}
\label{sec:numerics}
We analyze the GW emission for several 2D simulations performed
with the general relativistic neutrino hydrodynamics code
\textsc{Vertex-CoCoNuT} \citep{mueller_10} or its (pseudo)-Newtonian
counterpart \textsc{Vertex-Prometheus} \citep{rampp_02,buras_06_a}.
The reader should refer to these papers for an in-depth description of
the numerical methods and the input physics (including the choice of
progenitor models, neutrino interaction rates, and the equation of
state). At this point, we confine ourselves to a summary of those
technical aspects that are directly relevant for the present paper
with its focus on GW signals.

The hydrodynamics modules \textsc{CoCoNuT} and \textsc{Prometheus}
used in conjunction with the neutrino transport solver \textsc{Vertex}
both rely on similar higher-order finite-volume techniques, but differ
in their treatment of general relativity. \textsc{Prometheus} solves
the equation of Newtonian hydrodynamics, but can optionally account
for some effects of strong-field gravity by means of a modified
gravitational potential (``effective potential'';
\citealp{rampp_02,marek_06}). By contrast, the general relativistic
equations of hydrodynamics are solved in \textsc{CoCoNuT}
\citep{dimmelmeier_02_a,dimmelmeier_04}, and the extended conformal
flatness approximation (xCFC, \citealp{cordero_09}) is used for the
space-time metric. Comparisons of xCFC with the full ADM formalism in
the context of rotational core-collapse have shown excellent agreement
\citep{ott_06_a,ott_06_b,dimmelmeier_07_b} and suggest that it is
fully adequate for capturing the dynamics in the supernova core.
However, xCFC is a waveless approximation so that we need to extract
the GW signal in a post-processing step with the help of a suitable
version of the quadrupole formula (see Section~\ref{sec:extraction}).

The neutrino transport module \textsc{Vertex} solves the
energy-dependent neutrino moment equations for all neutrino flavors
using a variable Eddington factor technique \citep{rampp_02} and
relies on the so-called ``ray-by-ray-plus'' approximation for
multi-dimensional neutrino transport \citep{buras_06_a}. The
ray-by-ray-plus approach allows us to predict angular variations in
the neutrino radiation field (and hence the low-frequency GW signal
generated by the neutrinos) at least in rough qualitative agreement
with full multi-angle transport \citep{ott_08_a,brandt_10}.

In total, we simulate the evolution of six different progenitor models
with zero-age main sequence masses ranging from $8.1 M_\odot$ to $27
M_\odot$ (simulations G8.1--G27). These include a metal-poor $8.1
M_\odot$ ($10^{-4}$ solar metallicity) progenitor (model u8.1 in
\citealt{mueller_12b}), and a metal-free $9.6 M_\odot$ star (z9.6,
Alexander Heger, private communication), while the other progenitors
(models s11.2, s25.0 and s27.0 of \citealt{woosley_02} and model
s15s7b2 of \citealt{woosley_95}) have solar metallicity. Three of the
progenitors (z9.6, s25.0, s27.0) were simulated using the equation of
state of \citet{lattimer_91} with a bulk incompressibility modulus of
nuclear matter of $K=220 \ \mathrm{MeV}$ (LS220), whereas $K=180
\ \mathrm{MeV}$ (LS180) was applied in all other cases. Because of
very similar proto-neutron star radii for the neutron star masses
encountered in this study, LS220 and LS180 yield very similar results,
and the use of LS180 for some of the progenitors is justified despite
its marginal inconsistency with the $1.97 M_\odot$ pulsar
\citep{demorest_10}. Dynamical 1D simulations for the two
  EoSs with \textsc{Vertex} show a very similar evolution of the shock
  and proto-neutron star radius as well as the neutrino luminosities
  and mean energies (L.~H\"udepohl, private communication), a finding
  that is in agreement 
with 1D results of \citet{myra_94}, \citet{thompson_03}, and of
\citet{suwa_12}, who have also
  found an extremely similar behavior also in 2D.

The general features of the gravitational wave signal from the
post-bounce accretion and explosion phases are discussed on the basis of
the $11.2 M_\odot$ progenitor of \citet{woosley_02} and the $15
M_\odot$ model s15s7b2 of \citet{woosley_95} as prototypes for
``early'' and ``late'' explosions.  For the $15 M_\odot$ progenitor,
we also consider three models with a different treatment
of gravity (G15: GR; M15: effective potential; N15: purely Newtonian),
as well as a GR model (S15) with simplified neutrino rates
in order to explore the impact of these factors on the GW
  signal. The simplifications in model S15 include the use of the FFN
rates \citep{fuller_82,bruenn_85} for electron captures on nuclei, a
simpler treatment of neutrino-nucleon reactions, and the omission of
neutrino-neutrino pair conversion (see paper~II for details).

Except for the $25 M_\odot$ case, explosions have been obtained for
all progenitors with GR and with the full set of neutrino
rates. Figure~\ref{fig:model_dynamics} provides a compact overview
over the relativistic (G-)series of simulations: Models G8.1 and G9.6
explode rather early and exhibit convective activity only on a
moderate level after the onset of the explosion.  The $11.2 M_\odot$
model G11.2 shows a much slower expansion of the shock and several
violent shock oscillations before the explosion takes off. Model G15
develops a very asymmetric explosion as late as $\sim 450
\ \mathrm{ms}$.  The more massive $25 M_\odot$ and $27 M_\odot$ models
G25 and G27 differ from the other models by a more clearly discernible
SASI activity, visible as strong periodic sloshing motions of the
shock in Figure~\ref{fig:model_dynamics}, which lead to an explosion
in the case of G27. We note that no explosion develops in the
simulations without GR and/or the full neutrino rates (M15, N15, S15).
  
A summary of all nine models considered in this paper is given in
Table~\ref{tab:model_setup}.  For a detailed
discussion of models G11.2 and G15, see \citet{mueller_12}, and
for details on G8.1 and G27, see \citet{mueller_12b}.

\begin{table*}
  \caption{Model setup
    \label{tab:model_setup}
  }
  \begin{center}
    \begin{tabular}{ccccccccc}
      \hline \hline 
      &             & neutrino  & treatment of       & simulated                  & angular    & explosion  & time of                    &\\
      model & progenitor  & opacities & relativity   & post-bounce time           & resolution & obtained   & explosion\tablenotemark{a} & EoS  \\
      \hline
      G8.1 & u8.1   & full set    & GR hydro + xCFC    &  $325 \ \mathrm{ms}$  & $1.4^\circ$ & yes        & $175 \ \mathrm{ms}$ & LS180 \\
      G9.6 & z9.6   & full set    & GR hydro + xCFC    &  $735 \ \mathrm{ms}$  & $1.4^\circ$ & yes        & $125 \ \mathrm{ms}$ & LS220 \\
      G11.2 & s11.2   & full set    & GR hydro + xCFC    & $950 \ \mathrm{ms}$  & $2.8^\circ$ & yes        & $213 \ \mathrm{ms}$ & LS180 \\
      G15 & s15s7b2 & full set    & GR hydro + xCFC    &  $775 \ \mathrm{ms}$  & $2.8^\circ$ & yes        & $569 \ \mathrm{ms}$ & LS180 \\ 
      S15 & s15s7b2 & reduced set & GR hydro + xCFC    &  $474 \ \mathrm{ms}$  & $2.8^\circ$ & no         & --- & LS180 \\
      M15 & s15s7b2 & full set    & Newtonian +
                                    modified potential &  $517 \ \mathrm{ms}$  & $2.8^\circ$ & no         & --- & LS180 \\ 
      N15 & s15s7b2 & full set    & Newtonian (purely) &  $525 \ \mathrm{ms}$  & $1.4^\circ$ & no         & --- & LS180 \\ 
      G25 & s25.0   & full set    & GR hydro + xCFC    &  $440 \ \mathrm{ms}$  & $1.4^\circ$ & no         & ---  & LS220 \\
      G27 & s27.0   & full set    & GR hydro + xCFC    &  $765 \ \mathrm{ms}$  & $1.4^\circ$ & yes        & $209 \ \mathrm{ms}$ & LS220  \\
      \hline \hline
    \end{tabular}
    \tablenotetext{1}{Defined as the point in time when the average shock radius $\langle r_\mathrm{sh} \rangle$ reaches $400 \ \mathrm{km}$.}
  \end{center}
\end{table*}

\section{Gravitational Wave Extraction}
\label{sec:extraction}
The xCFC approximation used in \textsc{Vertex-CoCoNuT} does not allow
for a direct calculation of gravitational waves as the corresponding
degrees of freedom in the metric are missing. We therefore need to
extract gravitational waves in a post-processing step with the help of
some variant of the Einstein quadrupole formula \citep{einstein}.
Modified versions of the Newtonian quadrupole formula (exploiting
ambiguities concerning the identification of Newtonian and
relativistic hydrodynamical variables) have been found to be
reasonably accurate even in the strong-field regime
\citep{shibata_03,nagar_07,cordero_12}. For the gauge used in
\textsc{Vertex-CoCoNuT} and the typical conditions in a supernova
core, it is possible to derive a modified version of the time-integrated
Newtonian quadrupole formula \citep{finn_89,finn_90,blanchet_90}
directly from the field equations (see Appendix~\ref{app:tqf}). Assuming
axisymmetry, we obtain the quadrupole amplitude $A_{20}^\mathrm{E2}$
in non-geometrized units for spherical polar coordinates as
\begin{eqnarray}
\label{eq:gw_formula}
A_{20}^{\mathrm{E}2}&=&
\frac{32 \pi^{3/2} G}{\sqrt{15} c^4}
\int 
\, \ud \theta \, \ud r 
\phi^6 r^3 \sin \theta
\\
\nonumber
&&
\left\{
\frac{\pd }{\pd t}
\left[S_r \left(3 \cos^2\theta -1 \right) + 3 r^{-1} S_\theta \sin\theta \cos\theta\right]
\right.
\\
\nonumber
&& 
\left.
-
\left[\dot{S}_{r,\nu}, \left(3 \cos^2\theta -1 \right) + 3 r^{-1} \dot{S}_{\theta,\nu} \sin\theta \cos\theta\right]
\right\}
.
\end{eqnarray}
Here, $\phi$ is the (dimensionless) conformal factor for
the three-metric in the CFC spacetime, and $S_i$ denotes the covariant
components of the relativistic three-momentum density (in
  non-geometrical units, i.e.\ $S_r$ is given in  $\mathrm{g} \ \mathrm{cm}^{-2}
  \ \mathrm{s}^{-1}$ and $S_\theta$ in $\mathrm{g} \ \mathrm{cm}^{-1}
  \ \mathrm{s}^{-1}$) in the $3+1$ formalism, which is given in terms
of the rest-mass density $\rho$, the specific internal energy
$\epsilon$, the pressure $P$, the Lorentz factor $W$, and the
covariant three-velocity components $v_i$
as
\begin{equation}
S_i=\rho (1+\epsilon/c^2+P/\rho { c^2}) W^2 v_i.
\end{equation}
$\dot{S}_{i,\nu}$ denotes the momentum source term for $S_i$ due to
neutrino interactions (which must be subtracted from $\pd S_i/\pd t$
as explained in Appendix~\ref{app:tqf}). In practice, these neutrino
source terms do not yield a significant contribution to the integral
in Equation~(\ref{eq:gw_formula}).

$A_{20}^\mathrm{E2}$ determines the dimensionless strain measured by
an observer at a distance $R$ and at an inclination angle $\Theta$
with respect to the z-axis (see, e.g., \citealt{mueller_97_b}), 
\begin{equation}
\label{eq:strain}
h=\frac{1}{8} \sqrt{\frac{15}{\pi}} \sin^2 \Theta
\frac{A_{20}^\mathrm{E2}}{R}.
\end{equation}
In the following, we will always assume the most optimistic case of an
observer located in the equatorial plane, i.e.\ $\sin^2 \Theta = 1$.
In addition to the gravitational wave signal from the matter, we
compute the gravitational wave signal due to anisotropic neutrino
emission using the Epstein formula \citep{epstein_78,mueller_97} for
the gravitational wave strain $h_\nu$,
\begin{equation}
h_\nu = \frac{2 G}{c^4 R} \int \limits_0^t L_\nu (t') \alpha_\nu (t')\, \ud t'.
\end{equation}
Here $L_\nu$ is the total angle-integrated neutrino energy flux, and
the anisotropy parameter $\alpha_\nu$ can be obtained as
\begin{equation}
\label{eq:alpha_nu}
\alpha_\nu =
\frac{1}{L_\nu} \int \pi \sin \theta \left(2 |\cos
\theta| - 1\right) \frac{\ud L_\nu}{\ud \Omega} \ud \Omega
\end{equation}
in axisymmetry \citep{kotake_07}. $h_\nu$ can be converted
into an amplitude $A_{20,\nu}^\mathrm{E2}$ by inverting Equation~(\ref{eq:strain}).

The energy $E_\mathrm{GW}$ radiated in gravitational waves can be
computed from $A_{20,\nu}^\mathrm{E2}$ as follows (see, e.g.,
\citealt{mueller_97_b}),
\begin{equation}
E_\mathrm{GW}=\frac{c^3}{32 \pi G} \int \left(\frac{\ud A_{20,\nu}^\mathrm{E2}}{\ud t}\right)^2 \ud t.
\end{equation}
We also calculate the spectral energy distribution $\ud E / \ud f$ of
the gravitational waves,
\begin{equation}
\frac{\ud E}{\ud f}=
\frac{c^3}{16 \pi G} 
\left(2 \pi f\right)^2
\left|
\int\limits_{-\infty}^\infty e^{-2\pi i f t} A_{20}^\mathrm{E2}(t) \, \ud t
\right|^2.
\label{eq:gw_spectrum}
 \end{equation}

Previous studies \citep{marek_08,murphy_09,mueller_e_12} have relied
on Fourier transform methods (such as the short-time Fourier
transform) for analyzing temporal variations in the frequency
structure of the signal. While the Fourier transform of the signal can
be related directly to the power radiated in gravitational waves, the
time-variable frequency structure of the signal can be captured more
sharply with wavelet transforms.

We therefore compute the wavelet transform $\chi(t,f)$ (expressed as a
function of time and frequency) of the gravitational wave signal using
the Morlet wavelet (see, e.g., Equations~1 and 2 and Table~1 of \citealt{torrence_98}) with wavenumber
$k=20$ and define a normalized power spectrum $\bar{\chi}(t,f)$:
\begin{equation}
\label{eq:wavelet}
\bar{\chi}(t,f)=
\frac{f \left| \chi(t,f)\right|^2} 
{\underset{f' \in \mathbb{R}^+}{\max} \left(f' \left| \chi(t'=t,f')\right|^2\right)}.
\end{equation}
In other words: At a given time $t$, $\bar{\chi}$ is obtained by
dividing the weighted wavelet transform $f \left| \chi(t,f)\right|^2$
by its maximum over all frequencies for that time slice.  The
frequency weighting has been chosen in correspondence with the energy
spectrum in Equation~(\ref{eq:gw_spectrum}).\footnote{A weighting
  factor $f$ instead of $f^2$ is used because of the usual
  scale-dependent (i.e.\ frequency-dependent) normalization of the
  wavelet transform. } We found that this weighting and normalization
procedure helps to reveal the frequency structure of the signal most
perspicuously.

\begin{figure*}
\plottwo{f2a.eps}{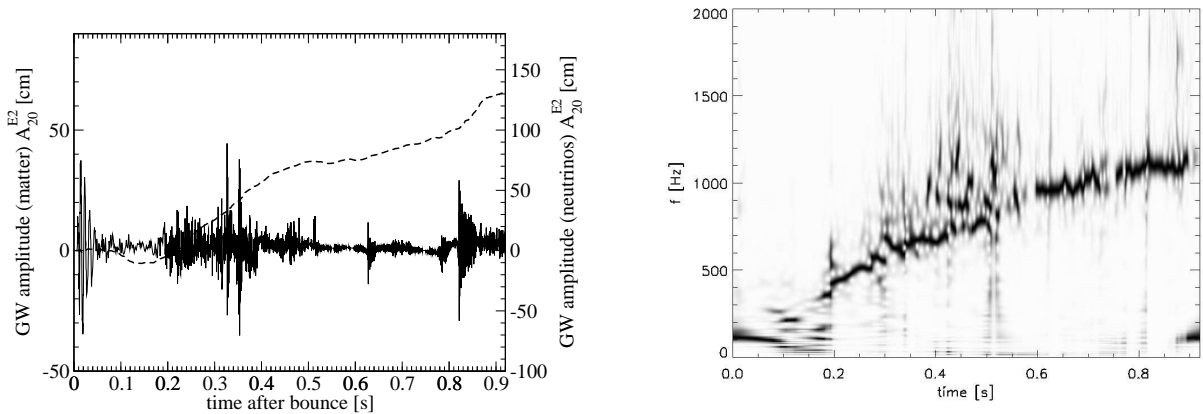}
\caption{Left panel: Matter (solid lines) and neutrino (dashed lines)
  gravitational wave signals for model G11.2. Note that the scale for
  the matter signal (left vertical axis) is different from the scale
  for the neutrino signal (right vertical axis). Right panel:
  Normalized wavelet spectrum $\bar{\chi}(t,f)$ of the matter signal
  for model G11.2. The grayscale ranges from white ($\bar{\chi}=0$) to
  black ($\bar{\chi}=1$, maximum value).  For the definition of
  $\bar{\chi}(t,f)$, see Equation~(\ref{eq:wavelet}).
\label{fig:g11}
}
\end{figure*}

\begin{figure*}
\plottwo{f3a.eps}{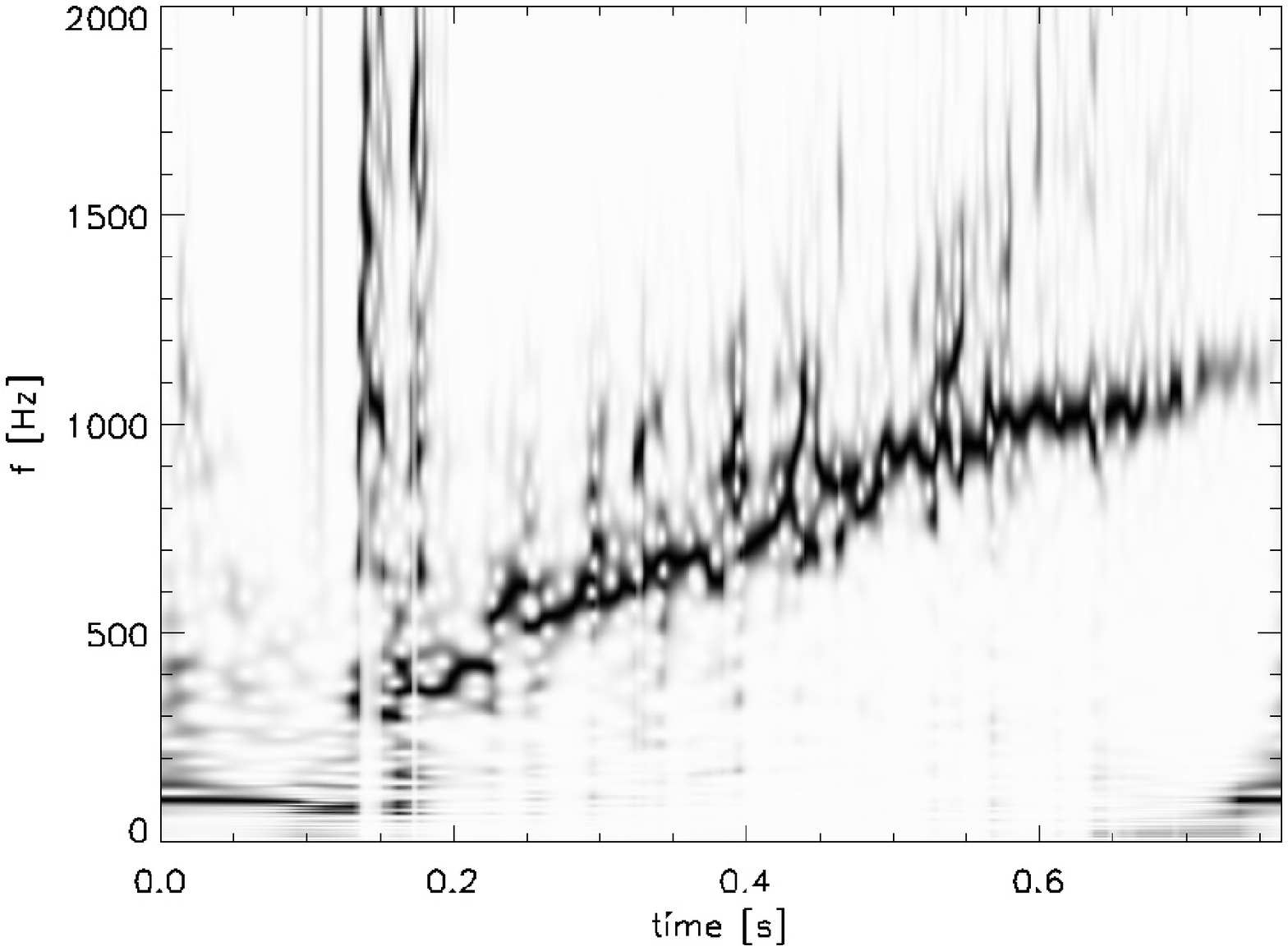}
\caption{Left panel: Matter (solid lines) and neutrino (dashed lines)
  gravitational wave signals for model G15. Note that the scale for
  the matter signal (left vertical axis) is different from the scale
  for the neutrino signal (right vertical axis). Right panel:
  Normalized wavelet spectrum $\bar{\chi}(t,f)$ of the matter signal
  for model G15. 
\label{fig:g15}
}
\end{figure*}

\begin{figure*}
\plottwo{f4a.eps}{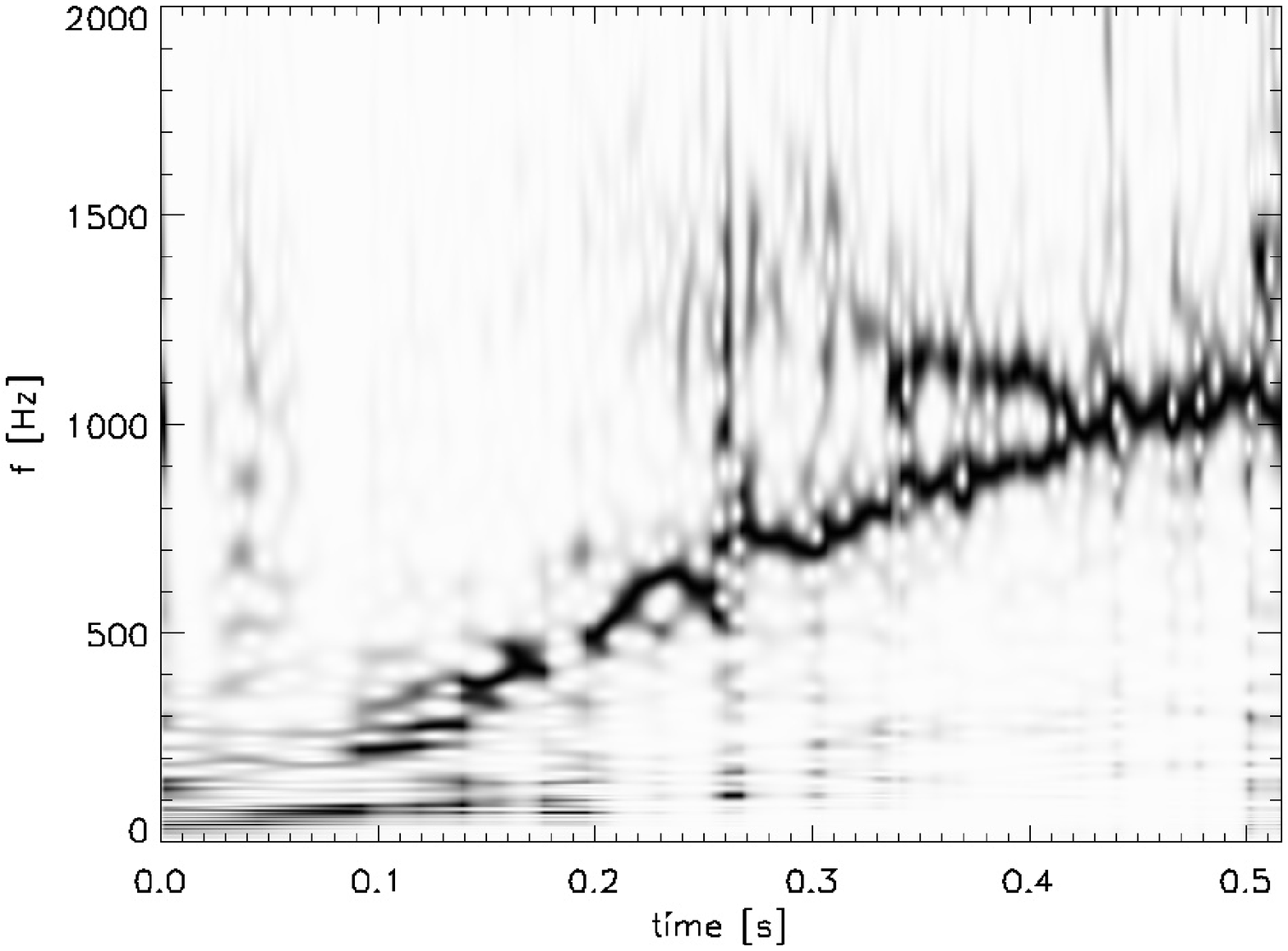}
\caption{Left panel: Matter (solid lines) and neutrino (dashed lines)
  gravitational wave signals for model M15. Right panel:
  Normalized wavelet spectrum $\bar{\chi}(t,f)$ of the matter signal
  for model M15. 
\label{fig:m15}
}
\end{figure*}

\begin{figure*}
\plottwo{f5a.eps}{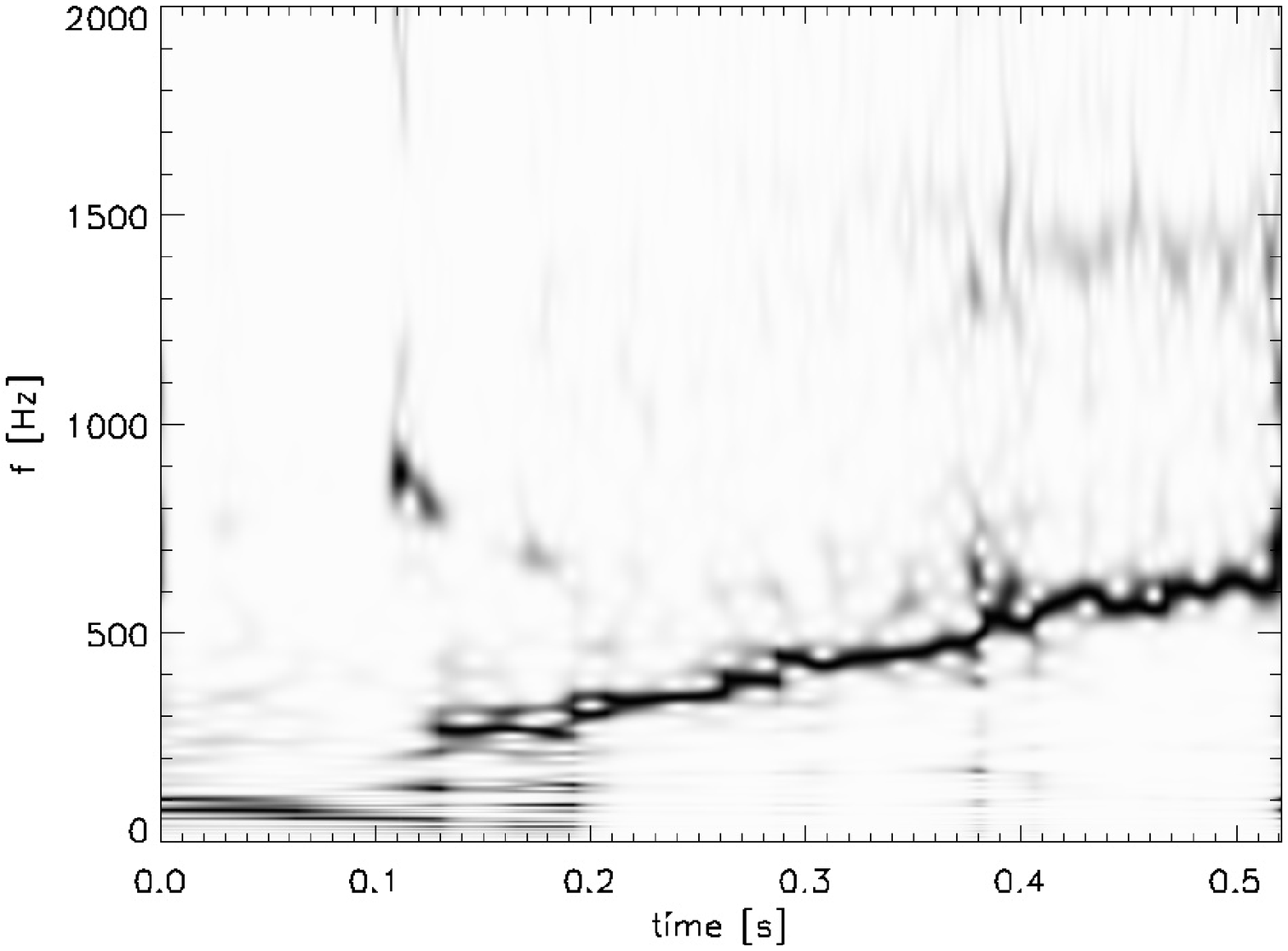}
\caption{Left panel: Matter (solid lines) and neutrino (dashed lines)
  gravitational wave signals for model N15. Right panel: Normalized
  wavelet spectrum $\bar{\chi}(t,f)$ of the matter signal for model
  N15.
\label{fig:n15}
}
\end{figure*}

\begin{figure*}
\plottwo{f6a.eps}{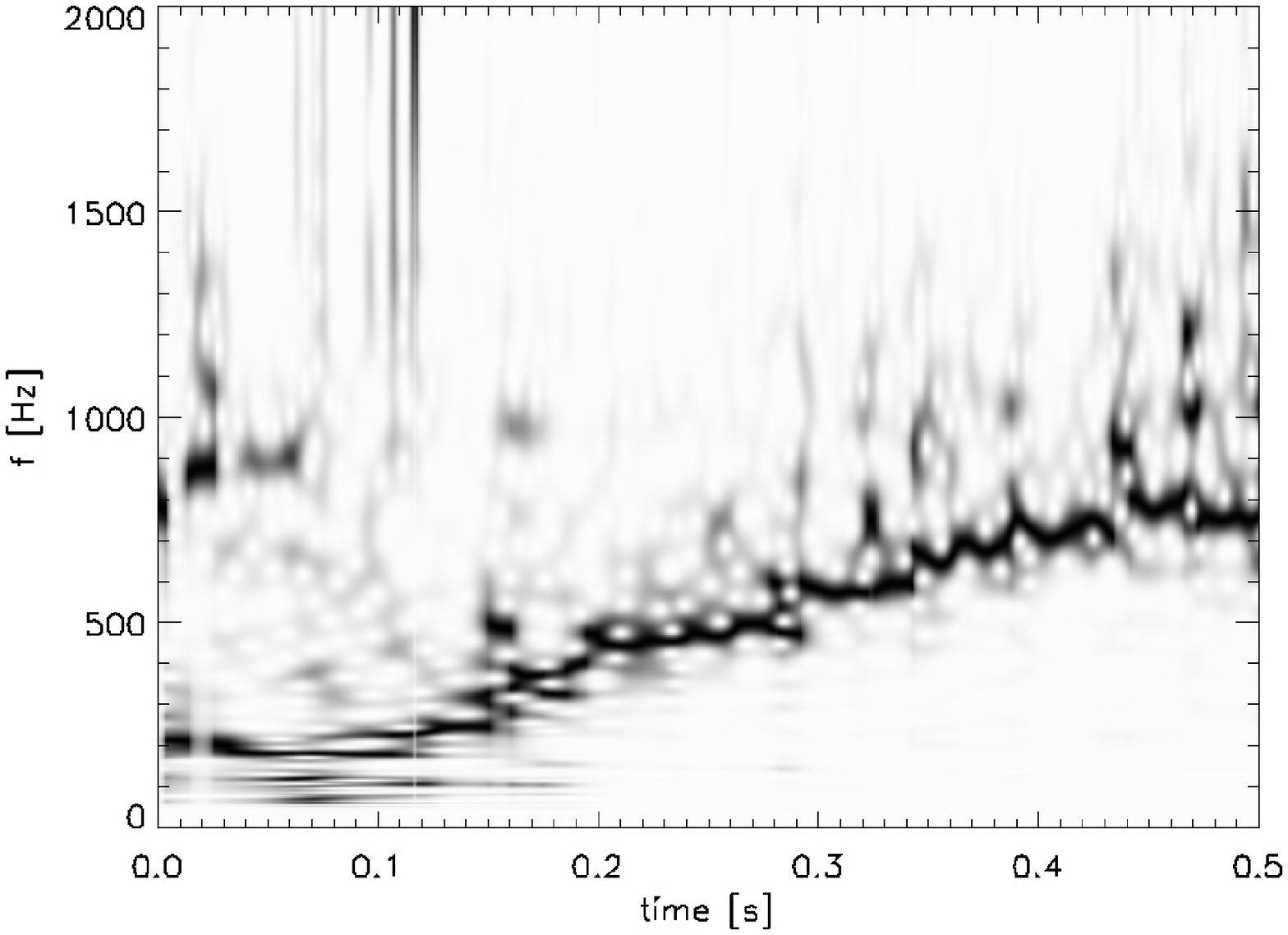}
\caption{Left panel: Matter (solid lines) and neutrino (dashed lines)
  gravitational wave signals for model S15. Note that the scale for
  the matter signal (left vertical axis) is different from the scale
  for the neutrino signal (right vertical axis). Right panel:
  Normalized wavelet spectrum $\bar{\chi}(t,f)$ of the matter signal
  for model S15. 
\label{fig:s15}
}
\end{figure*}

\section{Exemplary Discussion of GW Signal Features for
$11.2 M_\odot$ and $15 M_\odot$ Models}
\label{sec:general_features}

\subsection{Qualitative Description of the GW Signal}
GW amplitudes (both for the matter and neutrino signals) for models
G11.2, G15, M15, N15, and S15 are given in the left panels of
Figures~\ref{fig:g11}--\ref{fig:s15}, and the evolution of the signal
in frequency space is illustrated with the help of wavelet spectra in
the right panels of these Figures.  Qualitatively, our gravitational
waveforms share almost all the characteristics of waveforms recently
obtained from (pseudo-)Newtonian simulations of core-collapse
supernova explosions \citep{marek_08,murphy_09,yakunin_10}: We observe
a quasi-periodic signal during the phase of prompt post-shock
convection roughly between $10 \ \mathrm{ms}$ and $50 - 70
\ \mathrm{ms}$ after bounce, which is followed by a more quiescent
phase of several tens of milliseconds until hot-bubble convection and
increased SASI activity set in and produce a strong stochastic signal
component.  The stochastic signal in Figure~\ref{fig:g11} is strongest
during the first $200 \ \mathrm{ms}$ after the onset of the explosion,
and then continues at a lower amplitude with proto-neutron star
convection taking over as the dominant source for high-frequency
gravitational waves. Model G15 (Figure~\ref{fig:g15}) also shows a
``tail'' with a steadily increasing wave amplitude in the explosion
phase, a feature which is due to the expansion of a strongly prolate
shock as recognized by \citet{murphy_09} and \citet{yakunin_10}. In
addition, anisotropic neutrino emission gives rises to an almost
monotonically growing amplitude from about $200 \ \mathrm{ms}$ onward,
which is, in fact, much larger than the ``tail'' in the matter signal.
Model G11.2, on the other hand, exhibits a new feature: Here we observe
several bursts of high-frequency gravitational wave emission in the
explosion phase with amplitudes comparable to the phase of hot-bubble
convection -- a phenomenon that turns out to be connected to the
fallback of material onto the proto-neutron star in a rather weak
explosion \citep{mueller_12}.

The wavelet spectra in Figures~\ref{fig:g11}--\ref{fig:s15} equally
reflect the evolution through these different phases.  The early
quasi-periodic signal initially produces a peak at $\sim 100
\ \mathrm{Hz}$, which is clearly discernible during the first $\sim 100
\ \mathrm{ms}$. During the subsequent phase of reduced gravitational
wave activity, there is no clearly identifiable narrow emission band,
instead broadband low-frequency (model G11.2 until $\sim 200
\ \mathrm{ms}$) or high-frequency (model G15 around $\sim 150 \ \mathrm{ms}$) noise dominates the
spectrum. For models M15, N15, and S15 this intermediate phase is not
very pronounced in the wavelet spectra. The subsequent phase of strong
SASI activity and hot-bubble convection is mostly characterized by a
relatively narrow emission band, which appears between $100
\ \mathrm{ms}$ and $200 \ \mathrm{ms}$ (depending on the simulation)
and gradually shifts to higher frequencies at later times. Except for
model N15, there is also considerable gravitational wave activity at
frequencies above this band. This contribution disappears in the later
explosion phase ($\gtrsim 600 \ \mathrm{ms}$), leaving a single,
relatively sharply defined peak frequency. Note that low-frequency
components (i.e.\ asymmetric shock expansion) are poorly reflected in
the wavelet spectrum due to their small weighting factor (see
Equation~(\ref{eq:gw_spectrum})).

In the following, we shall discuss specific features of the individual
phases of gravitational wave emission in our simulations and study the
quantitative impact of GR on the GW signal in more detail. A cursory
glance at Figures~\ref{fig:g11}--\ref{fig:s15} already reveals that the
spectra contain ``cleaner'' information about the specific signal
properties of the different models than the stochastically varying
amplitudes, but we will nonetheless examine the signals extensively
both in the time and in the frequency domain.

\begin{figure}
\plotone{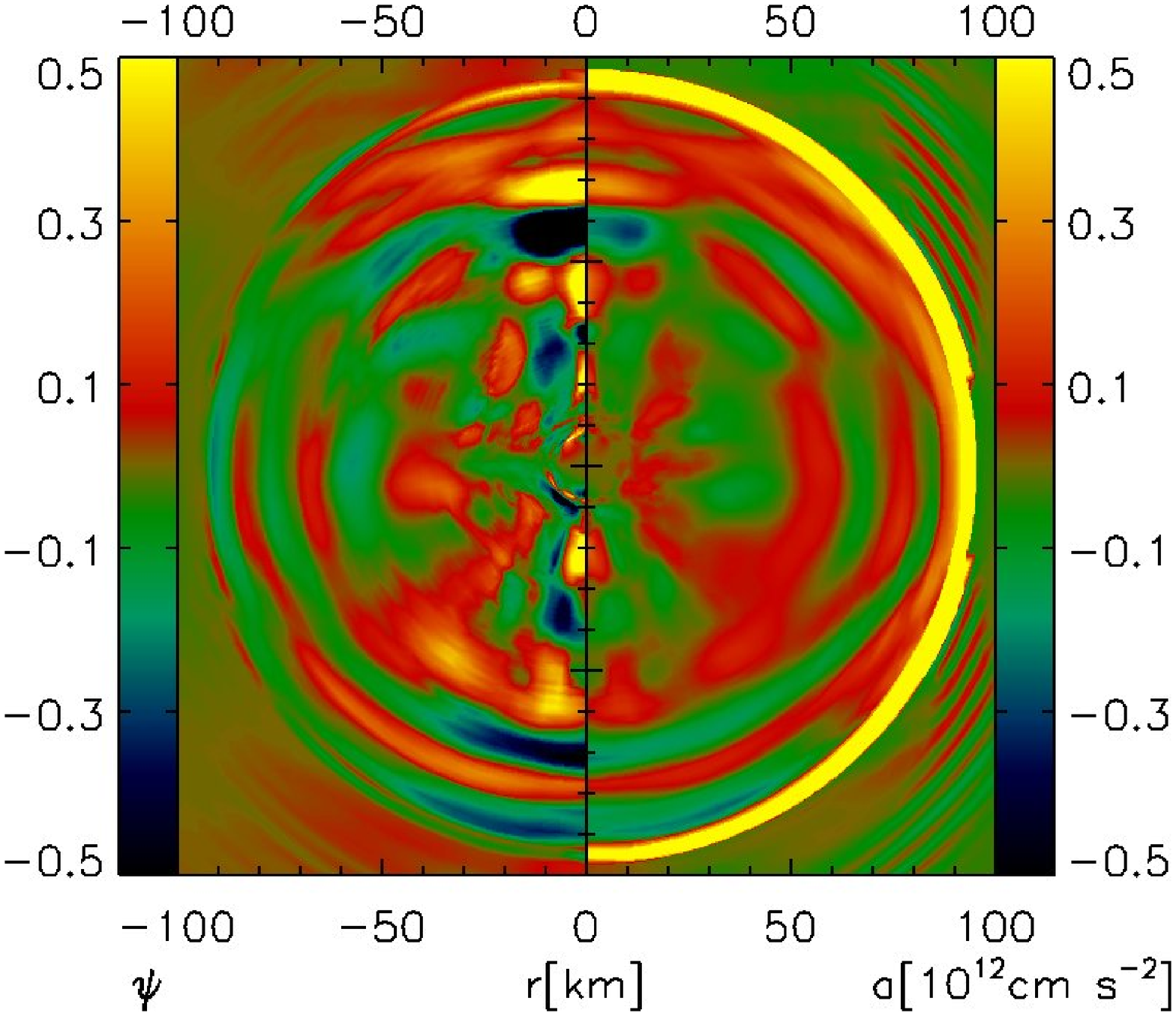}
\caption{The dimensionless integrand $\psi$ in the quadrupole formula
  (\ref{eq:gw_formula}) for the matter signal (left half of figure) and the time derivative $a=\pd
  v_r/\pd t$ (right half) of the radial velocity field $22 \ \mathrm{ms}$ after
  bounce for model G15.
\label{fig:acoustic_waves}
}
\end{figure}

\begin{figure}[t]
\plotone{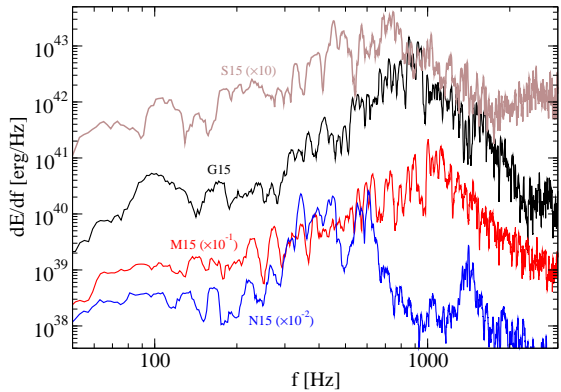}
\caption{
  Gravitational wave energy spectra (matter signal only) for models
  G15 (black), M15 (red) N15 (blue), and S15 (brown) for the time
  interval from $20 \ \mathrm{ms}$ to $520 \ \mathrm{ms}$ after
  bounce. The Fourier transform has been carried out without a window
  function in order to retain the high-frequency contribution from the
  phase of strong gravitational wave emission after $400
  \ \mathrm{ms}$ in model G15.  In order to better differentiate the
  curves, the spectra have been rescaled by a factor of $10^{-1}$,
  $10^{-2}$, and $10$ for model M15, N15, and S15, respectively.
  \label{fig:spectra_comparison}
}
\end{figure}

\begin{figure}
\plotone{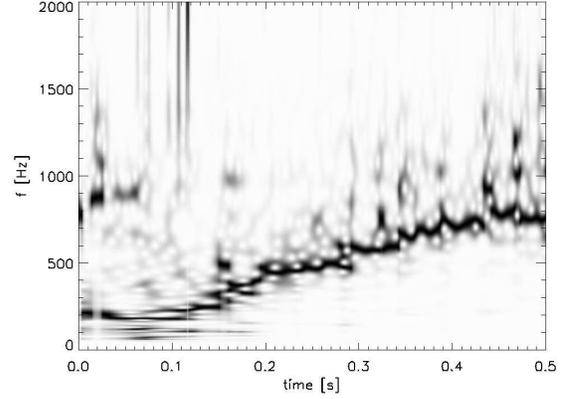}
\caption{ Maximum Brunt-V\"ais\"al\"a frequency $f_p$ in the
  convectively stable region between the proto-neutron star convection
  zone and the gain layer as a function of time for models G15 (solid
  line), M15 (dashed) and N15 (dotted). $f_p$ is a rather flat
  function of radius in this region, and the maximum value may
  therefore be taken as an estimate for the ``typical'' value of the
  Brunt-V\"ais\"al\"a frequency. The plot also shows the peak
  frequencies extracted from the wavelet spectra
  (Figures~\ref{fig:g15}, \ref{fig:m15}, \ref{fig:n15}) at intervals of
  $\approx 50 \ \mathrm{ms}$ (squares: model G15, circles: M15,
  triangles: N15).
\label{fig:dominant_frequencies}
}
\end{figure}

\subsection{Early Quasi-periodic Signal}
With regard to the early quasi-periodic signal, two major questions
ought to be addressed, namely whether GR affects the typical amplitude
and frequency of this signal component, and, even more basically, what is the
hydrodynamical origin of the GW signal during this phase? Different
authors have ascribed the quasi-periodic signal during the first
several tens of milliseconds directly to prompt post-shock convective overturn
\citep{marek_08,murphy_09}, while others \citep{yakunin_10} have
emphasized the contribution of early SASI motions. \citet{yakunin_10},
in particular, argued that the quasi-periodic signal is largely due to
the deceleration of the infalling matter at an oscillating aspherical
shock.

\subsubsection{Origin of the Signal}
\label{sec:origin_prompt_signal}
For our simulations, the question about the origin of the ``prompt
convection'' signal can be answered relatively clearly.  The
deceleration of matter at the shock can be ruled out as major
contribution factor with the help of an analytic estimate of the
expected (matter) gravitational wave amplitude
$A_{20,\mathrm{shock}}^\mathrm{E2}$ from a non-stationary aspherical
shock wave. For weak SASI oscillations, we obtain the following
formula in terms of the pre-shock density $\rho_\mathrm{p}$, the ratio
of the post-shock and pre-shock densities $\beta$, the power-law index
of the pre-shock density profile $\gamma$, and the multipoles $a_\ell$
of the angle-dependent shock position,
\begin{equation}
\label{eq:gw_shock_main}
  A_{20,\mathrm{shock}}^\mathrm{E2} \approx
  \frac{256 \pi^{3/2} G}{5 \sqrt{15} c^4}
  \rho_\mathrm{p}
  \left(\beta -1 \right)
  a_0^3 \left[(4+\gamma) a_2 \dot{a}_0+\dot{a}_2 a_0\right],
\end{equation}
as derived in Appendix~\ref{app:gw_shock}. The multipole coefficients
$a_\ell$ are defined in terms of the angle-dependent shock position
and the $\ell$-th Legendre polynomial as
\begin{equation}
a_\ell=
\frac{2 \ell + 1}{2}
\int\limits_{0}^\pi r_\mathrm{sh}(\theta) P_\ell(\theta) \, \ud \cos \theta.
\end{equation}
With typical values for the
early post-bounce phase ($\rho_\mathrm{p}\approx 10^{9} \ \mathrm{g} 
\ \mathrm{cm}^{-3}$, $\gamma \approx -1.5$), the resulting values are more
than an order of magnitude smaller than the observed amplitudes on the
order of several $10 \ \mathrm{cm}$.  

However, convection can be excluded as the direct source of the
quasi-periodic signal as well, since the destabilizing entropy and
lepton number gradients are erased very quickly, while the initial phase
of gravitational wave emission lasts for several tens of
milliseconds. The actual source of the signal can be determined by
considering the integrand
\begin{equation}
  \psi=
  r^3 \sin \theta
  \frac{\pd}{\pd t}
  \left[S_r (3 \cos^2 \theta - 1) + 3 r^{-1} S_\theta \, \sin \theta \, \cos
  \theta \right] \phi^6
\end{equation}
in the quadrupole formula (\ref{eq:gw_formula}) for the matter signal
in order to identify the main contributions to $A_{20}^\mathrm{E2}$
(cp.\ \citealp{murphy_09}). By visualizing $\psi$ (left half of
Figure~\ref{fig:acoustic_waves}), waves or hydrodynamic mass motions
responsible for gravitational wave emission can be identified fairly
well. The largest contribution to $A_{20}^\mathrm{E2}$ comes from
coherent stripe-like patterns between $25 \ \mathrm{km}$ and the
shock. The underlying flow pattern appears to be dominated by
propagating wavefronts (and not convective plumes), which emerge even
more clearly when the partial time derivative $a=\pd v_r/\pd t$ of the
velocity field is plotted (right panel of
Figure~\ref{fig:acoustic_waves}). These can be identified as acoustic
waves by their propagation speed and by the fact that the temporal
variations $\delta P$ and $\delta \rho$ of the pressure and the
density obey the relation $\delta P/P \approx \Gamma \delta \rho/\rho$
(where $\Gamma$ is the adiabatic exponent).  The frequency of these
waves is of the order of $100 \ \mathrm{Hz}$, which also accords well
with the gravitational wave spectrum during this signal phase.
In our model, prompt
convection is thus only indirectly responsible for the quasi-periodic
signal, either by directly initiating acoustic waves, or by instigating
SASI activity, which also involves the propagation of acoustic waves
in the post-shock region with the SASI frequency.

Both outgoing and ingoing waves (arising from partial reflection at
the shock) appear to be present so that it is tempting to interpret
these waves as transient p-modes in the post-shock region. It cannot
be excluded, however, that the oscillatory motions in the post-shock
regions also involve other (e.g.\ vorticity) waves. Along with the
transitory nature of these motions, this precludes an unambiguous
identification of the typical GW frequency with a definite mode
frequency based on the present simulation data.

\subsubsection{Effect of the GR Treatment on the Signal}
The amplitude and frequency of the quasi-periodic signal varies
appreciably between the models investigated here.  For the $15
M_\odot$ models with a different treatment of gravity (G15, M15, N15),
the maximum amplitude ranges from $26 \ \mathrm{cm}$ (G15) to $<2
\mathrm{cm}$ in model M15, where hardly any gravitational wave
activity takes place during this phase. We believe, however, that
these differences in amplitude may simply stem from a stronger or
weaker excitation of $l=2$ shock oscillations and acoustic waves in
the different models.  The differences are therefore not indicative of
a systematic effect of the GR treatment on the strength of this
signal. This conclusion is also supported by the fact that the early
signal is much stronger in model M15LS-2D of \citet{marek_08} (which
only differs from model M15 by a higher angular resolution) than in
model M15. Overall, the amplitudes in the GR case appear to be of
similar magnitude as in Newtonian \citep{kotake_07,murphy_09} and
pseudo-Newtonian simulations \citep{marek_08,yakunin_10} with peak
amplitudes $A_{20}^\mathrm{E2}$ of $\sim 40 \ \mathrm{cm}$ (G11.2) and
$\sim 30 \ \mathrm{cm}$ (G15).

On the other hand, the typical signal frequency is less affected by
such stochastic variations, and is therefore a better indicator for
systematic differences, which are indeed observed: The gravitational
wave spectrum of the early signal peaks at a somewhat higher frequency
of $\approx 100 \ \mathrm{Hz}$ in the GR case (G11.2 and G15) compared
to $\approx 60 \ \mathrm{Hz}$ in the Newtonian case (N15). For model
M15, the early signal is rather weak, but there is still a peak in the
region around $\approx 70 \ \mathrm{Hz}$ in the wavelet spectrum,
which is in agreement with the result obtained by \citet{marek_08} for
their $15 M_\odot$ model with the EoS of \citet{lattimer_91}. There
thus appears to be a tendency towards higher frequencies in the GR
case. This frequency shift could be the result of a
different width and location of the region affected by prompt
post-shock convection (cp.\ \citealt{marek_08} for this line of
reasoning in the context of EoS effects on the gravitational wave
signal): While prompt convection develops in the region between
enclosed masses of $0.61 M_\odot$ and $0.77 M_\odot$ in model G15, the
corresponding range in model N15 is $0.66 M_\odot \ldots 0.83 M_\odot$
and $0.61 M_\odot \ldots 0.71 M_\odot$ in model M15 and in the
Lattimer~{\&}~Swesty run of \citet{marek_08}. Interestingly,
\cite{marek_08} found a similar shift towards higher frequencies with
the stiffer nuclear equation of state of \citet{hillebrandt_85}, for
which the spectrum of the early gravitational wave signal is
remarkably similar to that of model G15.

The early GW signal from model S15 with simplified neutrino rates also
shows a different frequency structure than that of G15. The dominant
frequency is initially rather high ($\sim 200 \ \mathrm{Hz}$), but
after $50 \ \mathrm{ms}$ post-bounce, lower frequencies also appear in
the spectrum. Again, the width of the layer affected by prompt
convection is probably a factor responsible for this difference: In
S15, there are two unstable regions, namely $0.77 M_\odot \ldots 0.88
M_\odot$ and $0.96 M_\odot \ldots 1.07 M_\odot$, i.e.\ the convective
region contains a significantly larger mass and is located further
outside than in G15. It is noteworthy that the simplified neutrino
rates change the entropy and lepton number profiles in the early
post-bounce phase so drastically that the dynamics of prompt convection and
the GW signal are altered significantly.

Especially during the later phases of the prompt signal
  ($20 \ldots 100 \ \mathrm{ms}$ after bounce ) when the (gravito-)
  acoustic waves propagate throughout the entire region between the
  PNS and the shock, the GW frequency may also be influenced by the
  sound crossing time-scale or (if a vortical-acoustic loop is
  involved in the shock oscillations) the advection time-scale for
  that region. This could account for the frequency shift in the
  Newtonian case. Here, the shock radius, and hence the sound crossing
  time-scale as well as the advection time-scale are significantly
  larger than in the GR case during the relevant phase, which suggests
  lower GW frequencies.

We refrain from a more quantitative analysis of the early
  GW signal frequencies here. While it is obvious that the width and
  location of the region of prompt convection as well
as the shock position and PNS radius affect the parameters
  relevant for oscillations involving trapped (gravito-)acoustic waves
  (and perhaps vorticity waves) inside an accretion shock, i.e. the
  local sound speed, the gravitational acceleration, and the radial
  velocity, there is no simple quantitative theory for the early GW
  frequencies (different from the GW signal from hot-bubble convection
discussed in the next section).

\subsection{GW Signal from Hot-Bubble Convection and the SASI}
\subsubsection{Model Comparison -- Shift of Characteristic
Frequencies} After the early quasi-periodic signal has subsided, a
phase of relatively weak GW emission ensues, and even when hot-bubble
convection starts, the GW wave activity may not increase immediately.
The onset of stronger GW emission from hot-bubble convection appears
to be strongly model-dependent with no clear connection to the
dynamics. In model G11.2, the explosion is already underway when
the GW amplitude starts to increase significantly $200 \ \mathrm{ms}$ after
bounce, and in the Newtonian model N15, there is a similarly long
delay. At the other end of the scale, we find model S15, where
there is no gap between the early quasi-periodic signal
and the signal from hot-bubble convection (which is helped
by the fact that the gain region already develops
at $\sim 60 \ \mathrm{ms}$ in this model).

Due to the stochastic nature of the signal, the amplitudes of the
different models should be compared with some caution. All models
exhibit amplitudes on the order of several $10 \ \mathrm{cm}$, which
is roughly consistent with earlier pseudo-Newtonian studies with
sophisticated neutrino transport
\citep{mueller_04,marek_08,yakunin_10}. Nevertheless, two general
trends emerge: As exemplified by models G11.2 and G15, the phase of
strongest gravitational wave emission begins when the shock starts to
expand again and lasts. This phase lasts for about $200
\ \mathrm{ms}$, after which point the stochastic signal largely
subsides. Furthermore, the gravitational wave amplitude is loosely
correlated with the mass in the gain region and the convective energy,
which is why the pessimistic, non-exploding model M15 is characterized
by somewhat smaller amplitudes than G15, a model with stronger
convection and a larger gain region
(cp.\ \citealt{mueller_12}).  For the same reason, model S15 with
simplified neutrino rates and less favorable heating conditions also
shows smaller amplitudes than model G15 in general.

As for the early quasi-periodic signal, the gravitational wave
spectra reveal the effect of general relativity much more clearly than
the amplitudes. A sizable frequency shift in GR compared to the
Newtonian and the effective potential approximation can be seen both
in the wavelet spectra in Figures~\ref{fig:g15}--\ref{fig:n15} as well as in the
time-integrated spectral energy distribution $\ud E/\ud f$ (see
Equation~\ref{eq:gw_spectrum}) for the first $500 \ \mathrm{ms}$ after bounce
shown in Figure~\ref{fig:spectra_comparison}. In the GR run, the peak is
clearly located at a higher frequency than in the purely Newtonian
case, but the peak frequency remains somewhat smaller than with the
effective potential approach. To quantify these differences, we
compute the median $f_M$ of the spectral energy distribution $\ud E/
\ud f$, which is implicitly defined by the condition
\begin{equation}
\int\limits_0^{f_M} \frac{\ud E}{\ud f} \ud f
=
\frac{1}{2}\int\limits_0^{\infty} \frac{\ud E}{\ud f} \ud f,
\end{equation}
i.e.\ $f_M$ is the frequency below which half of the total energy in
gravitational wave is radiated.  We obtain values of $f_M=920
\ \mathrm{Hz}$ for model G15, $1100 \ \mathrm{Hz}$ ($+20\%$) for model
M15, and $510 \ \mathrm{Hz}$ ($-44\%$) for the purely Newtonian model
N15. The same ordering is observed in the wavelet spectra
(Figure~\ref{fig:g15}--\ref{fig:n15}) throughout the simulation once hot-bubble
convection starts. As for the early quasi-periodic signal, general
relativistic effects therefore considerably affect the gravitational
wave spectrum, leading to significantly higher frequencies (by $\sim
80 \%$) than in the purely Newtonian case, while the effective
potential approach proves to be accurate to within $\sim 20 \%$. Again
the effects of general relativity prove to be of similar magnitude as
EoS effects (for which we refer to \citealt{marek_08}) and emerge as a major
factor in determining the gravitational wave spectrum.

Unlike for the early quasi-periodic signal, the effect of the
simplified neutrino rates is not very pronounced. For model S15, we
obtain a value of $f_M=840 \ \mathrm{Hz}$, which is somewhat lower
than for model G15.  At a given time, the dominant frequencies
are relatively similar for both models (Figures~\ref{fig:g15},
\ref{fig:s15}), so the crucial factor for the shift of the median
frequency must be the onset of the explosion in model G15: This
leads to enhanced GW emission after $400 \ \mathrm{ms}$ and hence a
higher weighting factor for late-time high-frequency emission than for
model S15. 

We note that models G11.2 and G15 show very similar trends
in their wavelet spectra, but refer the reader to Section~\ref{sec:progenitors}
for a more thorough discussion of the progenitor dependence.

\subsubsection{Origin of Frequency Shift in GR}
The dependence of the typical frequency of the gravitational wave
signal from hot-bubble convection and SASI activity can be understood by
considering the dominant emission mechanism during this
phase. \citet{marek_08} and \citet{murphy_09} found that anisotropic
mass motions in the convectively stable region above the proto-neutron
star surface actually account for the bulk of the signal (although
these motions are in turn instigated by convection and the SASI).
While \citet{murphy_09} suggested the deceleration of infalling
convective plums as a direct source for the gravitational wave signal,
\citet{marek_08} also mention surface g-mode oscillations excited by the
downflows as a source. Both mechanisms can work in tandem, and one
may speculate that the continuous narrow emission bands in
Figures~\ref{fig:g11}--\ref{fig:s15} are associated with a time-variable, but
relatively stable g-mode frequency, whereas the deceleration of plumes
with random frequency and penetration depth is responsible for the
more noisy part of the spectrum above this band.

In either case, an anisotropic, buoyancy-driven flow is responsible
for the emission of gravitational waves, and the typical angular
frequency of the signal is therefore approximately given by the
buoyancy or Brunt-V\"ais\"al\"a-frequency $N$ in the convectively
stable region between the neutrinosphere and the gain radius (as
\citealt{murphy_09} explicitly verified for their model). As we shall
demonstrate, the GR treatment systematically affects the buoyancy
frequency in a way that accounts for the observed frequency
differences.

In the Newtonian case, the buoyancy frequency is given by the familiar
expression (see, e.g., \citealp{aerts}),
\begin{equation}
\label{eq:bv_newton}
N^2=\frac{1}{\rho}\frac{\pd \Phi}{\pd r} \left(\frac{1}{c_s^2}\frac{\pd P}{\pd r}-\frac{\pd \rho}{\pd r}\right),
\end{equation}
where $\Phi$ is the gravitational potential, $\rho$ the (rest-mass)
density, $c_s$ the sound speed, and $P$ the
pressure. Equation~(\ref{eq:bv_newton}) is not applicable in the
general relativistic case, however, because the underlying equations
of hydrodynamics are different. The relativistic expression for the
Brunt-V\"ais\"al\"a-frequency, derived in Appendix~\ref{app:bv_gr}
for the gauge used in \textsc{Vertex-CoCoNuT}, reads
\begin{equation}
\label{eq:bv_gr}
N^2=\frac{\pd \alpha c^2}{\pd r} \frac{\alpha}{\rho h \phi^4}\left(\frac{1}{c_s^2}\frac{\pd P}{\pd r}-\frac{\pd \rho (1+\epsilon/c^2)}{\pd r}\right),
\end{equation}
which contains correction terms involving the lapse function $\alpha$,
the conformal factor $\phi$, the specific internal energy $\epsilon$
and the specific relativistic enthalpy $h=1+\epsilon/c^2+P/(\rho
c^2)$.  Note that we formulated Equation~(\ref{eq:bv_gr}) in
terms of $\alpha$, $\phi$, and $h$ in their non-dimensional form.

In Figure~\ref{fig:dominant_frequencies}, we compare the real
frequency $f_p=N/2\pi$ (the ``plume frequency'' of
\citealt{murphy_09}) corresponding to the buoyancy frequency in the
convectively stable neutron star surface region for models G15, M15, and N15
and also show the peak frequency extracted from the wavelet spectrum
$\bar{\chi}$ at selected points in time. We clearly observe the same
ordering of the models for $f_p$ as for the peak frequency, although
$f_p$ generally overestimates the actual frequency of the spectrum by
up to $30\%$ (as already noticed by \citealt{murphy_09}).\footnote{The
  fact that the GW peak frequency is consistently lower than $f_p$ is
  consistent with the hypothesis that the dominant frequency seen in
  the GW spectrum is actually that of a surface g-mode oscillation, since the
  buoyancy frequency provides an upper bound for the g-mode frequency
  (see, e.g.,\citealp{kippenhahn,aerts}).}

The different buoyancy frequencies in G15, M15, and N15 thus account
nicely for the effect of the GR treatment on the GW spectrum, and the
terms responsible for the shift of $f_p$ can be readily identified: In
the Newtonian approximation (model N15), the neutron star is less
compact than in GR due to the lack of non-linear strong-field effects,
and the gravitational acceleration term in $\pd \Phi / \pd r$
Equation~(\ref{eq:bv_newton}) is therefore smaller than the
corresponding term $\pd \alpha c^2/\pd r$ in GR (or $\pd \Phi / \pd r$
in the effective potential approximation). $f_p$, and hence the GW
peak frequency is \emph{underestimated}.  The different compactness of
the proto-neutron star is most likely also responsible for the
discrepancy between our Newtonian results and those of
\citet{murphy_09}, who obtained even lower GW frequencies (e.g.\ only
\ $200 \ldots 300 \ \mathrm{Hz}$ at $500 \ \mathrm{ms}$): Their use of
a stiff EoS \citep{shen_98} and of a parametrized neutrino cooling and
heating scheme, which does not allow for the loss of energy and lepton
number from the PNS core, both tend to underestimate the contraction
of the proto-neutron star.

On the other hand, the effective potential approach (model M15) gives
a very good approximation for the compactness of the proto-neutron
star ($\pd \Phi/\pd r \approx \pd \alpha c^2/\pd r$), but still fails
to reproduce the correct peak frequency because the effects of relativistic
kinematics are not taken into account. In GR, the correction factor
$\alpha h^{-1} \phi^{-4}$ in Equation~(\ref{eq:bv_gr}) reduces $f_p$ by
about $15\%$ to $20\%$, which largely explains the lower frequencies
in model G15 compared to M15. In addition, the slightly higher
neutrino luminosities and mean energies in model G15 also correlate
with a
somewhat more shallow density gradient in the neutron star surface
region and thus reduce $f_p$ and the typical gravitational wave
frequency even a little further.

These arguments indicate that gravitational wave spectra from the
later ($\gtrsim 150 \ \mathrm{ms}$) post-bounce phase with an accuracy
better than $\sim 20\%$ in frequency space can only be obtained within
the framework of general relativistic hydrodynamics, since both the
Newtonian and the pseudo-Newtonian approximation suffer from an
\emph{intrinsic} accuracy limit.

\begin{figure*}
\plottwo{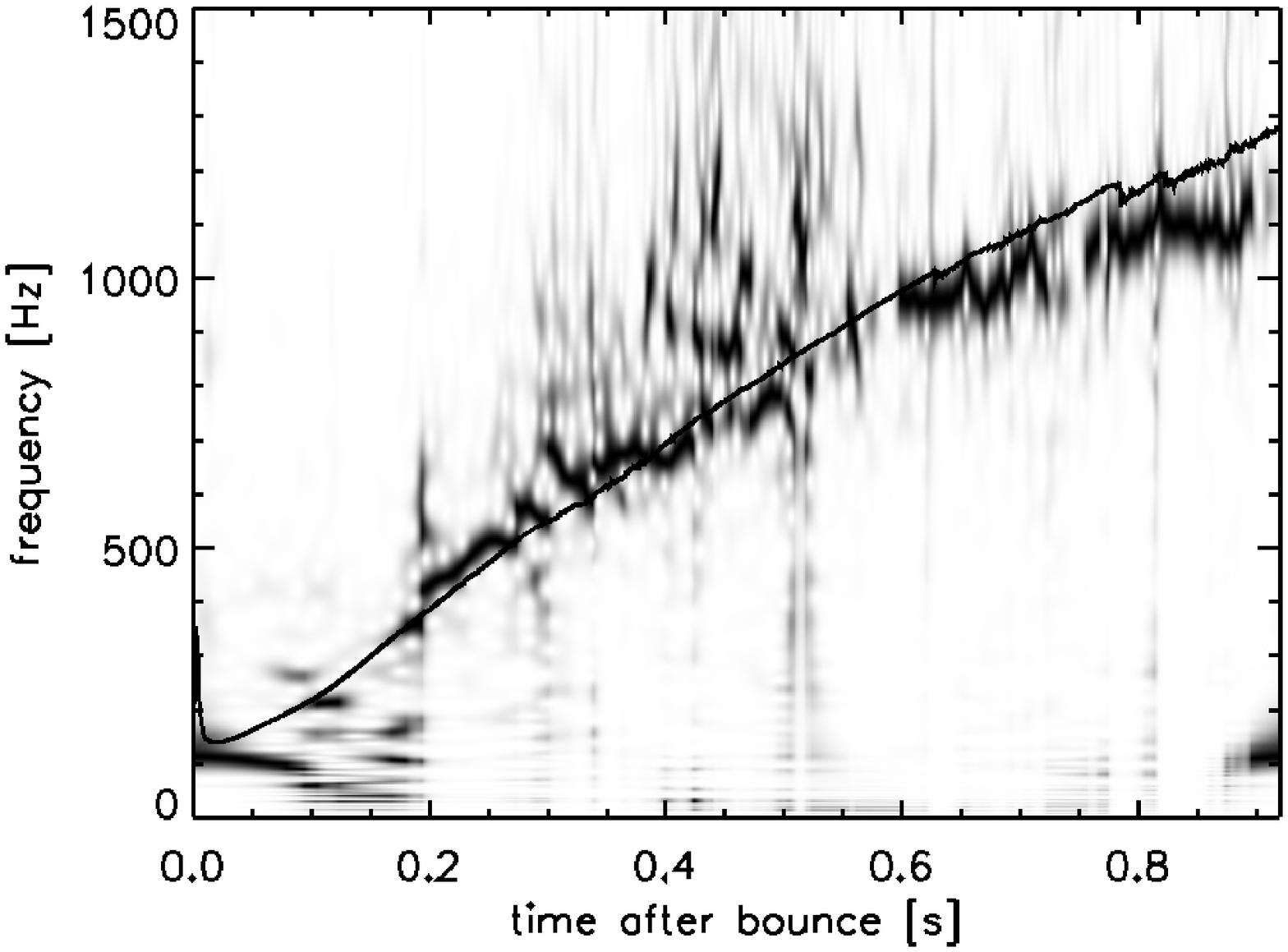}{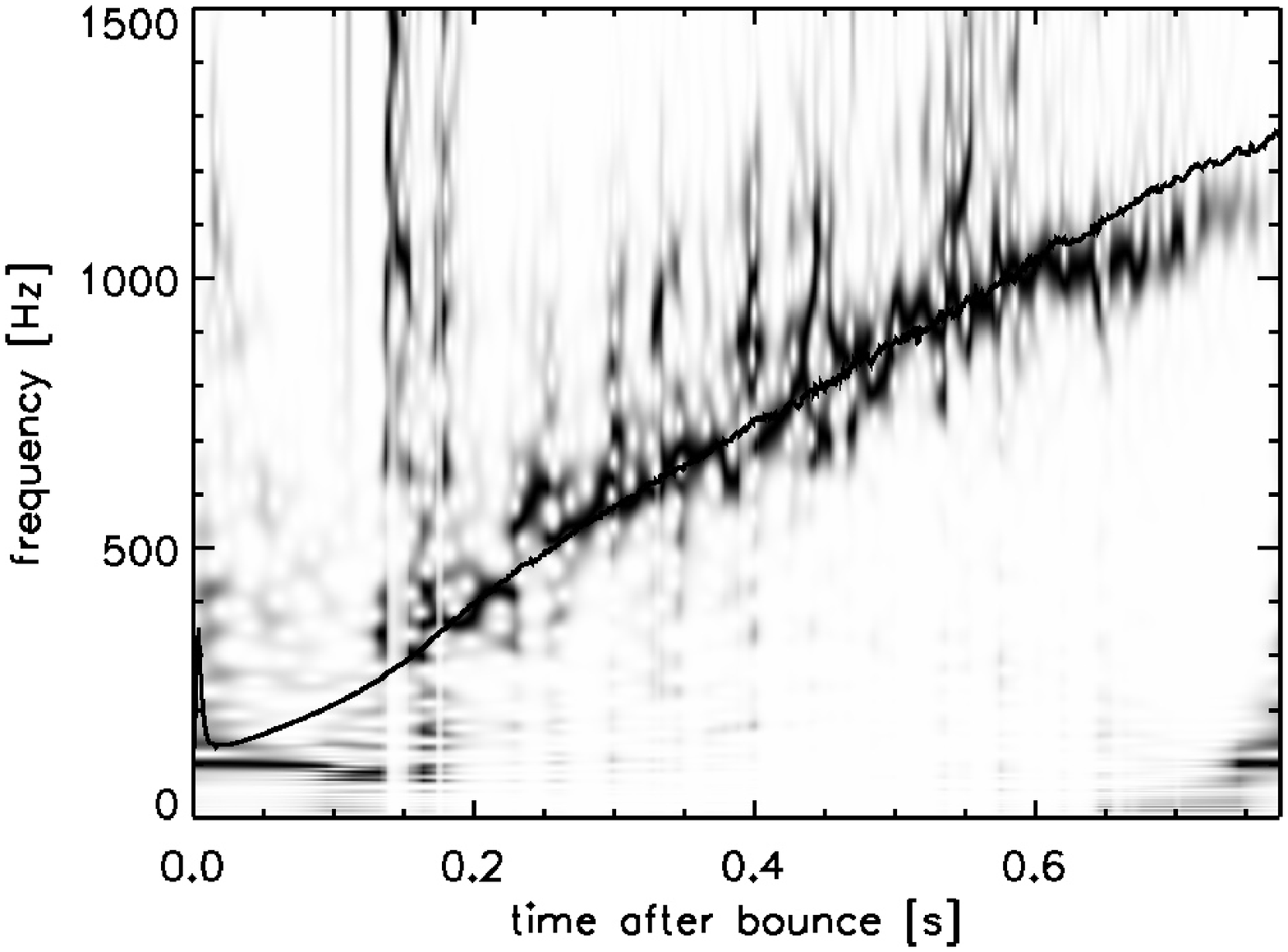}
\caption{Comparison of the wavelet spectra (identical to those in
  Figures~\ref{fig:g11} and \ref{fig:g15}, respectively) of model G11.2
  (left)and G15 (right) and the prediction of
  Equation~(\ref{eq:frequency_formula}) for the typical gravitational
  wave frequency in terms of the proto-neutron star mass and radius
  and the mean energy of electron antineutrinos measured by an
  observer at infinity.  Equation~(\ref{eq:frequency_formula})
  describes the evolution of the typical frequency fairly well and
  only overestimates $f_\mathrm{peak}$ somewhat at late times.
\label{fig:frequency_formula}
}
\end{figure*}

\subsubsection{Relation between PNS Properties and the Characteristic
GW Frequency}
\label{sec:f_peak}

Could the characteristic GW frequency, which emerges
clearly from the wavelet spectra in Figures~\ref{fig:g15}--\ref{fig:n15}
provide clues about properties of the proto-neutron star, such as its
mass, compactness, or surface temperature? A simple analytic estimate
for $f_p$ based on Equation~(\ref{eq:bv_gr}) can shed light on this question.

The metric functions $\alpha$ and $\phi$ in Equation~(\ref{eq:bv_gr})
can be approximated in terms of the proto-neutron star mass $M$ and
radius $R$ as $\alpha \approx \ln \alpha \approx \phi^{-2} \approx 1-G
M / (R c^2)$ at the proto-neutron star surface, and the pressure and
density gradients can be found by assuming a roughly isothermal
stratification (with temperature $T$) in the convectively stable
neutron star surface layer. Furthermore, an ideal gas equation of
state can be used as non-relativistic\footnote{This implies $\epsilon
  \ll 1$.} baryons dominate the pressure in the relevant region.  The
gradient terms in Equation~(\ref{eq:bv_gr}) can then be immediately
obtained in terms of $T$ and the neutron mass $m_n$ from the equation
of hydrostatic equilibrium,
\begin{equation}
\frac{\pd P}{\pd r} \approx
\frac{\pd}{\pd r}\left(\frac{\rho k T}{m_n}\right)=
-\rho \frac{\pd \ln \alpha}{\pd r}.
\end{equation}
Employing the relation $c_s^2=\Gamma k T/m_n$ for the speed of
sound ($\Gamma$ being the adiabatic index), we then arrive
at the following expression for
$f_p$,
\begin{equation}
f_p
=
\frac{N}{2 \pi}
=
\frac{1}{2\pi}
\frac{G M}{R^2} \sqrt{\frac{(\Gamma-1) m_n}{\Gamma k_b T}}
\left(1-\frac{G M}{R c^2}\right)^{3/2},
\end{equation}
where $\Gamma$ is the adiabatic index in the proto-neutron star surface region, and
$m_n$ is the neutron mass. The mean energy $\langle
E_{\bar{\nu}_e}\rangle$ of electron antineutrinos may be used as a
proxy for the temperature (with additional redshift corrections, i.e.\ 
$\langle E_{\bar{\nu}_e}\rangle \approx 3.151 \alpha(R) T$), and one thus
obtains a fairly accurate formula for the evolution of
the dominant GW frequency $f_\mathrm{peak}$:
\begin{equation}
\label{eq:frequency_formula}
f_\mathrm{peak}
\approx 
\frac{1}{2\pi}
\frac{G M}{R^2} \sqrt{1.1 \frac{m_n}{\langle E_{\bar{\nu}_e}\rangle}}
\left(1-\frac{G M}{R c^2}\right)^{2}.
\end{equation}
For our two exemplary models G11.2 and G15,
Figure~\ref{fig:frequency_formula} shows that
Equation~(\ref{eq:frequency_formula}) is in fairly good agreement with
the measured frequencies. The critical parameters regulating
$f_\mathrm{peak}$ are thus i) the surface gravity, ii) the surface
temperature, and iii) the compactness parameter $GM/Rc^2$ of the
proto-neutron star. Since $f_\mathrm{peak}$ also depends on the
thermal properties of the PNS surface layer, gravitational waves from
the accretion phase probably cannot provide an unambiguous probe for
the bulk properties of the proto-neutron star (mass, radius).  On the
other hand, the theory underlying
Equation~(\ref{eq:frequency_formula}) suggests that 2D models already
capture the frequency structure of the GW signal well: With the
buoyancy frequency in the convectively stable region determining the
GW spectrum (probably via the frequency of the $\ell=2$ surface
g-mode) we expect the same dominant frequency in 3D in the absence of
rotation (because of the degeneracy of oscillation modes with
different $m$).

Equation~(\ref{eq:frequency_formula}) illustrates that
$f_\mathrm{peak}$ is sensitive to factors that affect the contraction
and thermal evolution of the proto-neutron star, such as the EoS and
the neutrino treatment.  The different neutrino rates in model S15 are
not critical in this respect since they affect the proto-neutron star
surface temperature only on a very moderate level (see paper~II).
More radical approximations in the neutrino treatment could
potentially have a sizable impact, however, provided that they change
the proto-neutron star surface temperature and the neutrino mean
energies considerably.  The neutrino treatment (multi-group variable
Eddington factor transport vs. parametrized heating and cooling) along
with the different equation of state may also partially account for
the differences of our Newtonian model N15 to the models of
\citet{murphy_09}.  For the dependence of $f_\mathrm{peak}$ on the
progenitor, we refer the reader to Section~\ref{sec:progenitors}.

\begin{figure}
\plotone{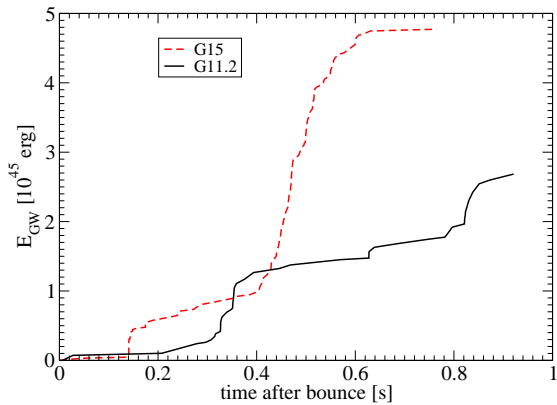}
\caption{Energy $E_\mathrm{GW}$ radiated in gravitational waves as
  function of time for the explosion models G11.2 (black solid line) and
  G15 (red dashed line) for the entire duration of the simulation. For
  model G15, most of the energy is radiated around the onset of the
  explosion between $400 \ \mathrm{ms}$ and $600 \ \mathrm{ms}$ after
  bounce.  By contrast, late-time GW bursts carry a sizable fraction
  of the radiated energy in the case of model G11.2.
  \label{fig:gw_energy}
}
\end{figure}

\begin{figure*}
\plotone{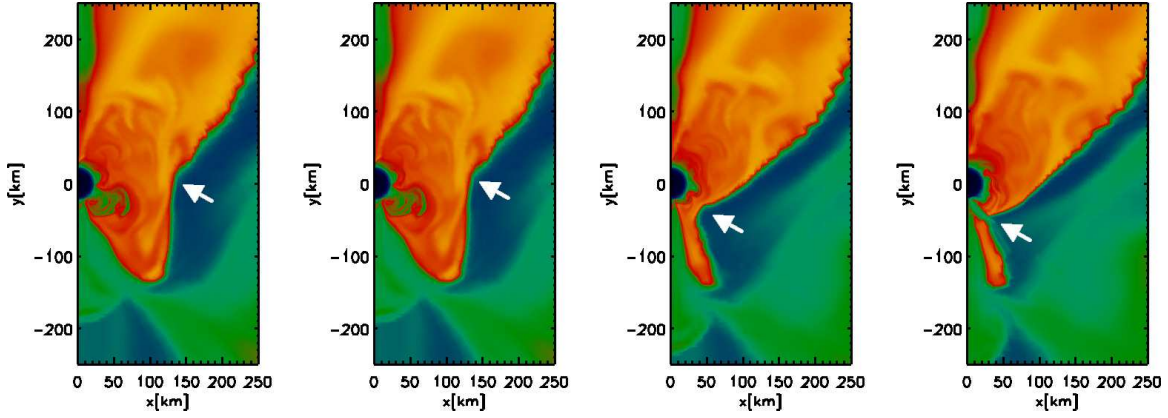}
\caption{Snapshots of the entropy $s$ in the region around the
  proto-neutron star at post-bounce times of $812.3 \ \mathrm{ms}$,
  $816.5 \ \mathrm{ms}$, $819.7 \ \mathrm{ms}$, and $821
  \ \mathrm{ms}$, depicting the formation of a new downflow by
  material falling through a hot bubble of neutrino-heated material.
  To guide the eye, the dent in the high-entropy bubble, which
  eventually develops into the new downflow is highlighted by a white
  arrow in each panel.  The values of the entropy range from
  $0k_b/\mathrm{nucleon}$ (black) to $35k_b/\mathrm{nucleon}$
  (yellow).  Due to its high infall velocity, the downflow overshoots
  far into the convectively stable proto-neutron star surface layer,
  exciting violent oscillations. The GW burst occurring at this time
  can be seen in Figures~\ref{fig:g11} and \ref{fig:correlation}.
\label{fig:downflow_burst}
}
\end{figure*}

\begin{figure}
\plotone{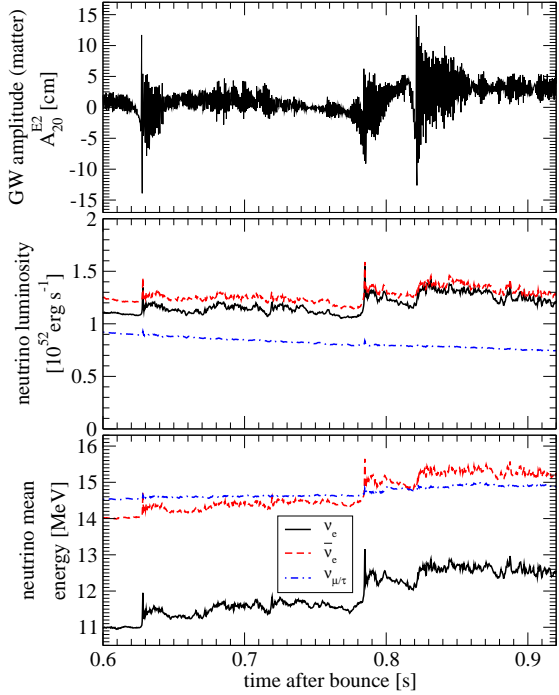}
\caption{Correlation of late-time bursts in the GW signal and the
  neutrino signal: The top panel shows the GW amplitude $A_{20}^{E2}$
  as a function of time for model G11.2 between $600 \ \mathrm{ms}$ and
  $920 \ \mathrm{ms}$ after bounce.  The middle panel displays the
  neutrino luminosity of $\nu_e$ (black solid line), $\bar{\nu}_e$
  (red, dashed), and $\nu_{\mu/\tau}$ (blue, dash-dotted) -- defined
  as the neutrino energy flux integrated over all emission directions
  as measured at an observer radius of $400 \ \mathrm{km}$. The bottom
  panel shows the direction-averaged mean energy of neutrinos at the
  same observer radius.
\label{fig:correlation}
}
\end{figure}

\begin{figure}
\plotone{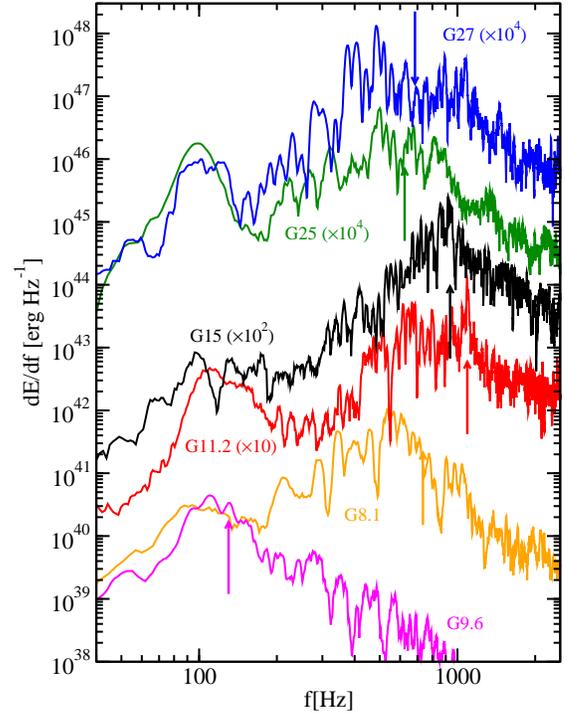}
\caption{Gravitational wave energy spectra (matter signal
  only) for the relativistic models G8.1 (orange), G9.6 (magenta), G11.2
  (red), G15 (black), G25 (green), and G27 (blue), for the entire
  duration of the simulation. Note that some of the spectra
  have been rescaled by a factor given along with the model
  designation. The Fourier transform has been carried
  out without a window function in order to retain the high-frequency
  contributions towards the end of some simulations. The spectra
  reflect larger amplitudes of the early quasi-periodic signal (around $\sim 100 \ \mathrm{Hz}$) in
  models G11.2, G25, and G27. For
  model G9.6, there is almost no high-frequency signal from hot-bubble
  convection. For G11.2, the sharp peak slightly above $1
  \ \mathrm{kHz}$ emerges in the spectrum due to late-time GW bursts.
  \label{fig:spectra_progenitors}
}
\end{figure}

\subsection{Explosion Phase -- Proto-Neutron Star Convection and Late-Time Bursts}
\label{sec:bursts}
As established by \citet{murphy_09} and \citet{yakunin_10}, the
emission of high-frequency gravitational waves subsides to a reduced
level once the shock accelerates outwards because the excitation of
oscillations in the proto-neutron star surface by violent convective
motions in the hot-bubble region largely ceases. This is well
reflected in the GW signals of models G11.2 and G15
(Figures~\ref{fig:g11}, \ref{fig:g15}) in the \emph{initial} phase
of the explosion, as the GW amplitude drops noticeably after $400
\ \mathrm{ms}$ and $600 \ \mathrm{ms}$, respectively. However, the
surface g-mode oscillations remain the dominant source of
high-frequency gravitational waves during this phase: The wavelet
spectra clearly show that the late-time signal comes from the same
emission band as the signal from the accretion phase (which shifts
continuously to higher frequencies as the neutron star contracts). An
analysis of the integrand $\psi$ in the quadrupole formula
(\ref{eq:gw_formula}) as in Section~\ref{sec:origin_prompt_signal}
also confirms that aspherical motions in the proto-neutron star
surface region are still the dominant source of gravitational
waves. Proto-neutron star convection now provides an excitation
mechanism for the oscillations
\citep{mueller_97,marek_08,murphy_09,mueller_e_12}, but due the
subsonic character (with Mach numbers $\lesssim 0.05$) of the
convective motions, the GW amplitude
$A_{20}^\mathrm{E2}$ remains rather small.

However, model G11.2 contradicts the established picture of subsiding GW
emission during the explosion. For this progenitor, we observe several
``bursts'' of stronger gravitational wave emission later in the
explosion phase on at least three occasions ($630 \ \mathrm{ms}$, $790
\ \mathrm{ms}$, and $820 \ \mathrm{ms}$), during which the amplitude
can become comparable to that coming from the phase of hot-bubble
convection. About $40\%$ of the GW energy is emitted later than $600
\ \mathrm{ms}$ after bounce (Figure~\ref{fig:gw_energy}), which is in
stark contrast to the more vigorous explosion in model G15 with faster
shock expansion.  These bursts occur when matter falls back onto the
proto-neutron star through a newly developing downflow, which excites
g-mode oscillations in the proto-neutron star surface region
(Figure~\ref{fig:downflow_burst}). The formation of new downflows in
the explosion phase is a consequence of the small explosion energy of
model G11.2 (see paper~II), and one could speculate that low-energy
fallback supernovae may generally reveal themselves through such
multiple GW burst episodes. While the development of such
  an accretion downflow may be facilitated by the constraint of
  axisymmetry, it is conceivable that funnel- or plume-shaped
  downdrafts may impinge on the proto-neutron star in a similar manner
  in a weak explosion in 3D (cf.\ some 3D results of
  \citet{mueller_e_12}, where long-lasting accretion after the
  explosion was associated with time-dependent accretion
  downdrafts.). Given the fact that \citet{takiwaki_12} do not obtain
  faster shock propagation in 3D than in 2D for this particular
  progenitor in their simulations using the IDSA scheme, it is not
  unlikely that the explosion energies will remain low as in our 2D
  model (a few $10^{49} \ \mathrm{erg}$, see \citealp{mueller_12}) and
  that late-time bursts seen in our simulations will survive in 3D.
  Without detailed 3D simulations of this sort, however, it
  is unclear how massive the downdrafts can become and what GW burst
  amplitudes they can cause.

Several interesting features of these late-time bursts should be
noted: Between the few instances where new accretion downflow develop,
aspherical mass motions above the proto-neutron star surface are
apparently not strong enough to produce significant noise in the GW
signal, and the g-mode frequency therefore emerges much more cleanly
from the spectrum during the late time GW bursts than in the accretion
phase. The narrow-band character of the spectrum
(see~Figure~\ref{fig:g11} and the red curve in
Figure~\ref{fig:spectra_progenitors}) is potentially helpful both for
the detection of the GW signal as well as for the interpretation of
the data (e.g.\ by allowing for a clearer discrimination of different
equations of state on the basis of the g-mode frequency). Moreover,
the bursts directly coincide with an enhancement of the electron
neutrino and antineutrino luminosities and mean energies as shown in
Figure~\ref{fig:correlation}: When the newly-formed downflows bring
fresh material into the cooling region, the luminosity of $\nu_e$ and
$\bar{\nu}_e$ increases abruptly by up to $\sim 20\%$, and the mean
energy of the emitted neutrinos jumps by up to $\sim 1
\ \mathrm{MeV}$. Such correlations between the neutrino and GW signals
probably merit further investigation, but a deeper analysis taking
into account both the observer position and the detailed anisotropies
in the neutrino radiation field in the manner of \citet{mueller_e_12}
will not be attempted here. The anisotropic neutrino emission will be
discussed more thoroughly in a subsequent publication.

\subsection{The Tails in the Matter and Neutrino Signals}
At late times, the gravitational wave signals of our models G11.2 and G15
exhibit the same low-frequency features that have been observed in
previous studies: The matter signal develops an offset or ``tail''
which has been ascribed to aspherical shock propagation
\citep{murphy_09,yakunin_10}, with prolate/oblate explosions leading
to positive/negative amplitudes. This offset is of the order of $10 -
15 \ \mathrm{cm}$ for model G15 (which is consistent with its prolate
explosion geometry), while only a small offset of $\approx 3
\ \mathrm{cm}$ develops for model G11.2 at rather late times due to the
small shock deformation. These amplitudes are consistent with a
direct numerical evaluation of
Equation~(\ref{eq:gw_shock_exact})\footnote{Note that
  Equations~(\ref{eq:gw_shock_main},\ref{eq:gw_shock}) cannot be used
  to evaluate the gravitational wave amplitude due to aspherical shock
  propagation for model G15 because the Legendre coefficient $a_1$ of
  the angle-dependent shock position $r(\theta)$ is of the same order
  as $a_0$, and the assumptions used to derive
  Equations~(\ref{eq:gw_shock_main},\ref{eq:gw_shock}) therefore break
  down.}, which confirms that aspherical shock propagation is indeed
responsible for the offset in the signal.

However, as in \citet{yakunin_10} the ``tail'' signal is much weaker
than the similar tail-like signal from the anisotropic emission of
neutrinos. Models G11.2 and G15 both develop long-lived polar downflows
that subsist well into the explosion phase and lead to a sustained
quadrupolar emission anisotropy over several hundreds of milliseconds;
in model G15 the running average of the anisotropy parameter
$\alpha_\nu$ (Equation~\ref{eq:alpha_nu}) over $50 \ \mathrm{ms}$ can
become as large as $0.02$. This is in striking contrast to model M15
with many alternating episodes of enhanced emission from the polar and
the equatorial region. It is very likely, however, that stochastic
model variations affect the neutrino-generated gravitational wave
signal considerably; model M15, e.g., is distinguished from the L\&S
model of \citet{marek_08} only by a different angular resolution, yet
\citet{marek_08} obtained a large negative signal amplitude due to the
predominant emission of neutrinos from the equatorial region.
However, even in model M15, the typical amplitude of the
neutrino-generated signal is larger than the typical signal from
aspherical shock propagation, which will probably always remain a
subdominant contribution to the low-frequency part of the spectrum.

\begin{figure*}
\plotone{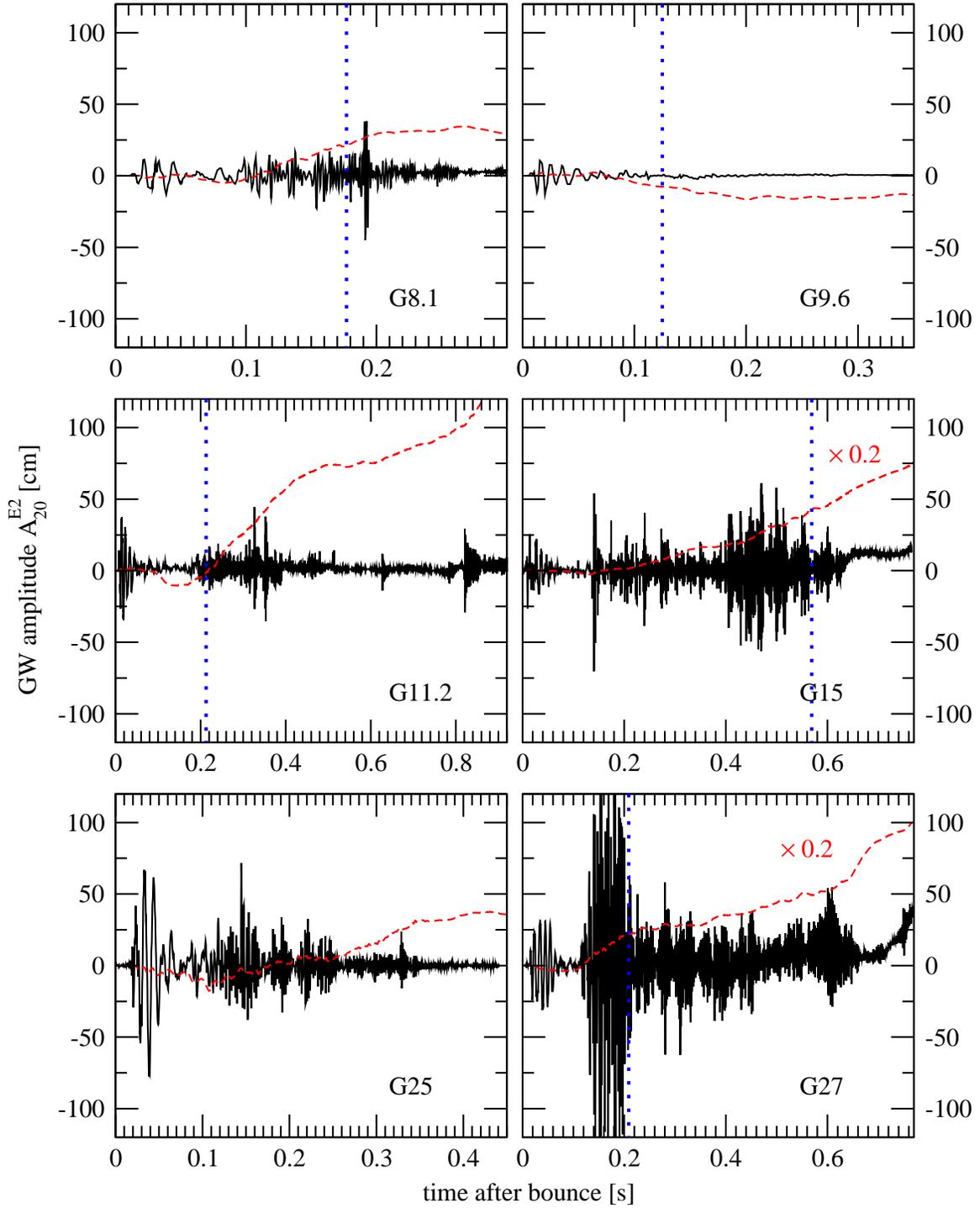}
\caption{Matter and neutrino gravitational wave amplitudes
  $A_{20}^{\mathrm{E2}}$ from GR simulations of the six progenitors
  considered in this paper. The matter and neutrino signals are shown
  as solid black and dashed red curves, respectively. For models G15
  and G27, the gravitational wave signal from the anisotropic emission
  of neutrinos has been rescaled by a factor of $0.2$. For the
  exploding models, the onset of the explosion (defined as the time
  when the average shock radius reaches $400 \ \mathrm{km}$) is
  indicated by a dotted blue line.
\label{fig:panorama}
}
\end{figure*}

\begin{table*}
  \caption{Progenitor dependence of the GW signal
    \label{tab:progenitors}
  }
  \begin{center}
    \begin{tabular}{ccccccccccc}
      \hline \hline
      Model & $\dot{M}_{90}$\tablenotemark{a}& $E_\mathrm{GW}$\tablenotemark{b} & $A_{20,max}^\mathrm{E2}$ \tablenotemark{c} & $f_{250}$\tablenotemark{d} & $f_{300}$\tablenotemark{e} & $f_{400}$\tablenotemark{f} & $f_M$\tablenotemark{g} & $M_\mathrm{PNS}$\tablenotemark{h} & $t_\mathrm{expl}$\tablenotemark{i} &\\
            & $(M_\odot \ \mathrm{s}^{-1})$ & $(10^{45}\ \mathrm{erg})$ 
                             &   $(\mathrm{cm})$       &   $(\mathrm{cm})$       &   $(\mathrm{cm})$       &  $(\mathrm{Hz})$ & $(\mathrm{Hz})$ & $(M_\odot)$ & $(\mathrm{ms})$\\
      \hline
      G8.1  & $0.38$ & $0.36$     & $45$  & $540$ & $ 600$ & ---   & $730$   & $1.37$ & $177$ \\
      G9.6  & $0.09$ & $0.0032$   & $13$  & ---   & ---    & ---   & $130$   & $1.36$ & $125$ \\
      G11.2   & $0.57$ & $2.7$      & $52$  & $570$ & $ 580$ & $680$ & $1100$  & $1.35$ & $213$ \\
      G15   & $1.08$ & $4.8$      & $70$  & $520$ & $ 580$ & $710$ & $950$   & $1.58$ & $569$  \\ 
      G25   & $1.80$ & $1.2$      & $77$  & $660$ & $ 750$ & $900$ & $620$   & $2.01$ & --- \\
      G27   & $1.44$ & $20$       & $253$ & $580$ & $ 700$ & $860$ & $680$   & $1.77$ & $209$ \\
      \hline
    \end{tabular}
    \tablenotetext{1}{Mass accretion rate $90 \ \mathrm{ms}$ after bounce.}
    \tablenotetext{2}{Energy of radiated gravitational waves.}
    \tablenotetext{3}{Maximum GW amplitude of the matter signal.}
    \tablenotetext{4}{Dominant emission frequency $250 \ \mathrm{ms}$ after bounce.}
    \tablenotetext{5}{Dominant emission frequency $300 \ \mathrm{ms}$ after bounce.}
    \tablenotetext{6}{Dominant emission frequency $400 \ \mathrm{ms}$ after bounce.}
    \tablenotetext{7}{Median frequency of the GW energy spectrum.}
    \tablenotetext{8}{Baryonic mass of proto-neutron star by the end of the simulation.}
    \tablenotetext{9}{Explosion time, defined by the instance when the average shock radius $\langle r_\mathrm{sh} \rangle$ reaches $400 \ \mathrm{km}$.}
  \end{center}
\end{table*}

\begin{figure}
\plotone{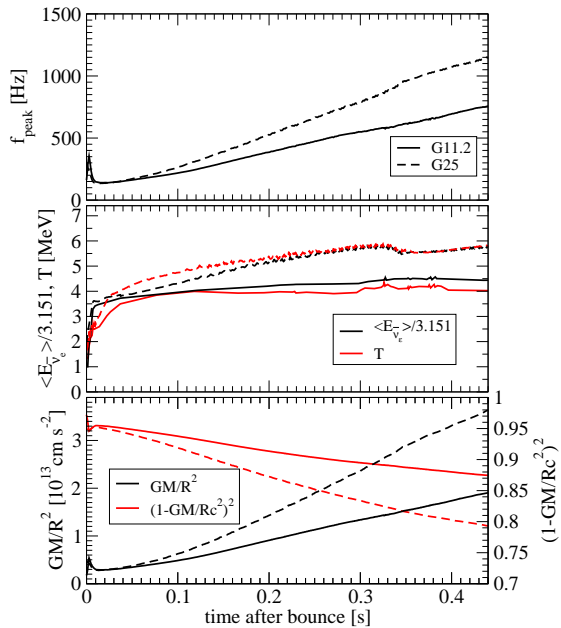}
\caption{Comparison of the terms contributing to the gravitational
  wave peak frequency $f_\mathrm{peak}$ in
  Equation~(\ref{eq:frequency_formula}) for models G11.2 (solid lines)
  and G25 (dashed lines).  The proto-neutron star surface
  temperature $T$ and the temperature $\langle E_{\bar{\nu}_e}/3.151
  \rangle$ corresponding to the mean electron antineutrino energy
  (assuming a Fermi distribution with vanishing chemical potential)
  are shown in the middle panel. The estimated neutron star surface
  gravity $G M/R^2$ and the the relativistic correction factor $(1-G
  M/Rc^2)^2$ in Equation~(\ref{eq:frequency_formula}) are shown in
  the bottom panel, and the resulting estimate for the peak frequency
  is depicted in the top panel. Note that we define the neutron star
  surface by a fiducial density of $10^{11} \ \mathrm{g}
  \ \mathrm{cm}^{-3}$.
\label{fig:progenitor_dependence}
}
\end{figure}

\section{Progenitor Dependence of the GW Signal}
\label{sec:progenitors}
After having discussed the different phases of GW emission for models
G11.2 and G15, we now turn to the progenitor dependence of the GW
signal. Models G8.1--G27 exhibit a widely different
  behavior in the accretion and (where applicable) the explosion
  phase, and thus illustrate how the GW emission is correlated with
  important parameters characterizing the dynamical evolution such as
  the explosion time, the proto-neutron star mass, or the mass
  accretion rate.  While the precise dynamics of individual
  progenitors (time and energy of the explosion, etc.) may depend on uncertainties in the
modeling (3D effects, progenitor structure, etc.), the spectrum of models
presented here already reveals such general trends very well.
 An overview over the waveforms for models
G8.1--G27 is given in Figure~\ref{fig:panorama}, and time-integrated
GW energy spectra are shown in Figure~\ref{fig:spectra_progenitors}.
Table~\ref{tab:progenitors} summarizes important characteristic
numbers for the waveforms and spectra.

\subsection{Progenitor Dependence of Waveforms}
Figure~\ref{fig:panorama} and Table~\ref{tab:progenitors} illustrate a
few general trends in the GW signals: GW activity tends to increase
for models with higher accretion rate $\dot{M}$ during the phase of
convective and SASI activity. Table~\ref{tab:progenitors} shows that
except for model G25 there is a strong correlation between the
accretion rate $\dot{M}_\mathrm{90}$ at a post-bounce time of $90
\ \mathrm{ms}$ (typically close to the onset of strong
convection/SASI) and the total gravitational wave energy
$E_\mathrm{GW}$ as well as the maximum amplitude
$A_{20,\mathrm{max}}^\mathrm{E2}$. This correlation is reasonable,
since the typical GW amplitude depends on the mass in the gain and
cooling regions participating in violent aspherical motions
and g-mode activity, respectively, Both of these masses
in turn regulated by the accretion rate. Naturally,
$\dot{M}_\mathrm{90}$ should not be understood as a single unambiguous
parameter regulating the GW emission; it is rather the overall evolution
of $\dot{M}$ that is relevant.

Because of vastly different accretion rates, models G9.6 and G27
constitute two extreme cases on the scale of GW emission: In model
G9.6, the accretion rate drops so quickly that there is almost no
signal from hot-bubble convection, and only the early quasi-periodic
signal remains. Moreover, although there is some convective overturn
in the ejecta in this model, the rather rapid explosion largely
precludes the excitation of PNS surface g-modes by
downflows or accretion downdrafts. The matter affected by overturn is blown
out rather quickly ahead of the developing neutrino-driven wind
(cp.\ top right panel of Figure~\ref{fig:model_dynamics}).

By contrast, a lot of mass is involved in strong SASI oscillations
prior to the onset of the explosion in model G27 (bottom right panel
of Figure~\ref{fig:model_dynamics}). Moreover, the shock does not
expand as rapidly afterwards, thus allowing violent aspherical motions
to continue for several hundreds of milliseconds beyond the onset of
the explosion. Consequently, the energy emitted into gravitational
waves is as high as $2 \times 10^{46} \ \mathrm{erg}$ for this model.

While the models G8.1, G11.2, and G15 fit nicely into a continuum
between these extreme cases, model G25 shows \emph{less} GW activity
than model G27 although it is characterized by even higher mass
accretion rates as well as by continuing SASI activity (bottom left
panel of Figure~\ref{fig:model_dynamics}). This reversal of the
aforementioned trend is related to fact that G25 does not develop an
explosion within the simulation time.  GW emission is typically
strongest around the onset of the explosion (G8.1, G11.2, G15, and G27),
when the mass in the gain region increases considerably and aspherical
motions in the post-shock region become most violent (i.e.\ reach the
highest velocities) before being quenched again during the later
phases of shock expansion. This phase is lacking in model G25, and the
strong retraction of the shock instead reduces the mass in the gain
region so that the GW amplitude is strongly attenuated during the
later evolution. Moreover, the relatively pure $\ell=1$ SASI mode seen
in model G25 may not be very efficient in exciting $\ell=2$
perturbations in the flow to which the GW amplitude is sensitive.

The overall trends in the progenitor dependence suggested by the six
models G8.1--G27 can thus be summarized as follows: GW emission tends
to become stronger with increasing mass accretion rate. In addition,
both a strong shock retraction and a very rapid explosion can quench
GW activity.

\subsection{Progenitor Dependence of GW Spectra}
To characterize the spectral properties of models G8.1--G27, we show
time-integrated energy spectra in
Figure~\ref{fig:spectra_progenitors}, for which we give the median
frequency $f_M$ in Table~\ref{tab:progenitors}.  Moreover, we list the
dominant emission frequencies $f_{250}$, $f_{300}$, and $f_{400}$ around
$250 \ \mathrm{ms}$, $300 \ \mathrm{ms}$, and $400 \ \mathrm{ms}$
after bounce to show the time-frequency evolution of the GW signals.

For the early quasi-periodic signal, which emerges as a low-frequency
peak in the curves of Figure~\ref{fig:spectra_progenitors}, we find no
significant progenitor dependence. This signal component always peaks
around $100 \ \mathrm{Hz}$ in our models. The later signal from
convection and the SASI shows somewhat larger variations.  The typical
emission frequencies $f_{250}$, $f_{300}$, and $f_{400}$ vary by $\sim
30 \%$, and tend to be highest for model G25. Considering that we
cover a very wide range in PNS masses from $\sim 1.36 M_\odot$
(G9.6, G11.2) to $\sim 2.0 M_\odot$ (G25), this variation may seem
astonishingly small. This can be understood as a partial cancellation
of terms in Equation~(\ref{eq:frequency_formula}) for
the dominant emission frequency: 
\begin{equation}
f_\mathrm{peak}
\approx 
\frac{1}{2\pi}
\frac{G M}{R^2} \sqrt{1.1 \frac{m_n}{\langle E_{\bar{\nu}_e}\rangle}}
\left(1-\frac{G M}{R c^2}\right)^{2}.
\end{equation}
While the surface gravity $G M/R^2$ is systematically larger for more
massive and compact neutron stars, the proto-neutron star surface
temperature is also higher, and the general relativistic correction
term $(1-G M/(R c^2))^{2}$ is smaller. The remaining net change is
typically moderate, as illustrated by G11.2 and G25 as rather extreme
examples in Figure~\ref{fig:progenitor_dependence}: Although the
surface gravity is higher by up to $\sim 90\%$ in model G25,
$f_\mathrm{peak}$ is only lower only by $\sim 35\%$ in model G11.2. Among
the less massive proto-neutron stars (models G8.1, G11.2, G15), the
differences are far less pronounced. The variation between
progenitors is similar to that found by \citet{murphy_09},
although the absolute values for the frequencies are
very different (since \citealt{murphy_09} lacked 
self-consistent neutrino transport and GR in their
simulations, and used different progenitors as well as a different
EoS).

When we consider time-integrated GW energy spectra
(Figure~\ref{fig:spectra_progenitors}), the relation between PNS and GW
properties becomes more complicated, however. We observe the highest
median frequency of $f_M=1100 \ \mathrm{Hz}$ for model G11.2, while we
find relatively low values between $600 \ \mathrm{Hz}$ and $700
\ \mathrm{Hz}$ for G25 and G27, although these two models have higher
values for $f_{250}$, $f_{300}$, and $f_{400}$. This reversal occurs
because much of the gravitational wave emission in G11.2 happens 
after the onset of the explosion in late-time bursts (see
Section~\ref{sec:bursts}). GW waves with relatively high frequencies
therefore contribute more strongly to the spectrum than in models such
as G25 and G27. These late-time bursts are distinctly visible as a
peak around $1100 \ \mathrm{Hz}$ for G11.2 in
Figure~\ref{fig:spectra_comparison}. For model G15, the situation is
somewhat similar, because the bulk of GW emission occurs around $500
\ \mathrm{ms}$ due to the rather late explosion. Model G9.6
constitutes another extreme because the high-frequency component
of the spectrum is largely absent, resulting in a low median
frequency of $f_M=130 \ \mathrm{Hz}$.

Our results demonstrate that time-integrated spectra are the result of
a somewhat complicated interplay between the time of strongest GW
emission and the time-dependence of the dominant emission
frequency. This implies that time-integrated spectra are even more
difficult to relate to PNS properties than $f_\mathrm{peak}$
(cp.\ Section~\ref{sec:f_peak}).

\section{Summary and Conclusions}
\label{sec:conclusions}
Based on several recent two-dimensional relativistic explosion models
\citep{mueller_12} for non-rotating progenitors between $8.1 M_\odot$
and $27 M_\odot$, we studied the GW signal of core-collapse supernovae
from the early post-bounce phase well into the explosion phase. We not
only provided gravitational waveforms from GR hydrodynamics
simulations with sophisticated multi-group neutrino transport for the
first time, but also analyzed the impact of the GR treatment (GR
vs.\ effective potential approach vs. Newtonian approximation) and of
the neutrino physics input on the GW signal by means of three
complementary (non-exploding) simulations of the $15 M_\odot$
progenitor.

In all phases, we found that GR has a sizable impact on the GW
spectrum. For purely Newtonian models, we found that the typical GW
frequency during the phase of hot-bubble convection is severely
underestimated (by $40\% \ldots 50\%$) compared to the GR case, while
it tends to be overestimated by $\sim 20 \%$ in the effective
potential approach. We determined that this is the results of
\emph{systematic} differences of the buoyancy frequency in the
proto-neutron star surface region (which approximates the typical GW frequency
as pointed out by \citealt{murphy_09}). In the Newtonian
approximation, the smaller compactness and surface gravity of the
neutron star lead to a lower buoyancy frequency compared to GR,
while the effective relativistic potential approach fails to correctly reproduce
the frequency of oscillations in the proto-neutron star surface
because of missing relativistic correction terms in the equations of
hydrodynamics. Similarly large differences as for the phase of
hot-bubble convection were also seen for the quasi-periodic signal
produced by prompt convection and early SASI activity.

Apart from systematic GR effects, we observed novel features in the GW
signal during the explosion phase for the $11.2 M_\odot$ progenitor, a
model that is likely to represents a fallback supernova.  Here the
fallback of material onto the proto-neutron star through newly-forming
downflows in the explosion phase leads to a strong excitation of
surface g-mode oscillations, which produce burst-like GW signals that
are sharply defined in frequency space. These late-time bursts are
correlated with small bursts of enhanced $\nu_e$ and $\bar{\nu}_e$
emission. Contrary to what more energetic explosion models
\citep{murphy_09,yakunin_10} might suggest, strong GW emission in the
high-frequency band late in the explosion phase appears to be possible
for underenergetic supernovae. In all other respects, however, our
models qualitatively confirm the four-phase picture of gravitational
emission with an early quasi-periodic signal, a quiescent phase of
several tens of milliseconds, a strong stochastic GW signal lasting
until some fraction of a second after the onset of the explosion, and
a low-frequency tail-signal from anisotropic neutrino emission and
shock expansion.

Our simulations also revealed trends for the progenitor dependence of
the GW signal. 
We found a variation of the maximum signal amplitude by
more than an order of magnitude between our two extreme cases with
$9.6 M_\odot$ and $27 M_\odot$. In general, high mass accretion rates
turn out to be conducive to strong GW activity provided that an
explosion still develops.  The time-dependent dominant emission
frequency varies by $\sim 30 \%$ with a tendency towards higher values
for more massive proto-neutron stars. The explosion dynamics plays a
major role for determining the time-integrated spectrum, which is
therefore more difficult to relate to PNS properties.

While the GW signal predictions presented here rely on
  state-of-the-art 2D supernova simulations with sophisticated
  neutrino transport, further research into the gravitational wave signals
  clearly remains a priority in core-collapse supernova studies.
  Model uncertainties, e.g. concerning the nuclear equation of state,
  the progenitor structure, or the role of 3D effects, need to be
  eliminated in order to predict precise amplitudes and frequencies
  for specific stellar progenitors. In particular, 3D
  modeling will be essential for determining the correct signal
  amplitudes for both the plus and cross polarization modes (the latter
  of which is not present in our models due to the assumption of
  axisymmetry)  simply because the energy contained in aspherical
  motions will be distributed differently among more available modes
  than in 2D.  If 3D effects change the overall dynamics of the
  accretion and explosion phase considerably, e.g.\ by allowing faster
  and more energetic explosions, this will also have a direct impact
  on the GW emission and may modify the details of the progenitor
  dependence.  However, the present controversy about the role of 3D
  effects (\citealp{nordhaus_10,hanke_12,burrows_12,murphy_12,dolence_12,couch_12,takiwaki_12}; see \citealt{janka_12b} for an overview discussion)
  precludes any premature statement. Different from the case of the GW
  amplitudes, we do not expect that 3D effects will change the
  frequency structure of the GW signal considerably.

At any rate, our findings clearly demonstrate that a proper treatment
of GR is indispensable for obtaining accurate predictions of
gravitational waveforms and spectra. In particular, GR simulations
will be required to link the GW signal to the properties of the
nuclear equation of state, or of the progenitor. However, as a
cautionary remark, we emphasize that the inclusion of GR effects at
the expense of detailed multi-group neutrino transport may potentially
yield little gain in our opinion. Although the simplifications of the
neutrino rates considered as test case in this paper cause a
significant difference only for the early quasi-periodic signal, more
radical approximations in the neutrino transport sector may
significantly affect the contraction, the compactness. and the surface
stratification of the neutron star, and might thus severely alter the
GW spectra. For quantitative predictions of GW signals,
future 3D simulations of core-collapse supernovae will thus have
to include both GR (cf.\ \citealp{kuroda_12}) and reliable neutrino transport.

\acknowledgements This work was supported by the Deutsche
Forschungsgemeinschaft through the Transregional Collaborative
Research Center SFB/TR 7 ``Gravitational Wave Astronomy'' and the
Cluster of Excellence EXC 153 ``Origin and Structure of the Universe''
(http://www.universe-cluster.de). The computations were performed on
the IBM p690 and the SGI Altix 3700 of the Computer Center Garching
(RZG), on the Curie supercomputer of the Grand \'Equipement National
de Calcul Intensif (GENCI) under PRACE grant RA0796, on the Cray XE6
and the NEC SX-8 at the HLRS in Stuttgart (within project SuperN), on
the JUROPA systems at the John von Neumann Institute for Computing
(NIC) in J\"ulich (through grant HMU092 and through a DECI-7 project
grant), and on the Itasca Cluster of the Minnesota Supercomputing
Institute.


\appendix

\section{Gravitational Wave Extraction: The Matter Signal}
\label{app:tqf}
Although the standard Newtonian quadrupole formula for the matter
signal and its time-integrated varieties
\citep{finn_89,finn_90,blanchet_90} as used in the \textsc{CoCoNuT}
code so far \citep{dimmelmeier_02_b,dimmelmeier_08} has proved
reasonably accurate (i.e.\ on a level of $10\% \ldots 20 \%$) even for
relatively strong gravitational fields in comparison with more
sophisticated direct wave extraction methods
\citep{shibata_03,nagar_07}, it suffers from several drawbacks: Its
application in the context of general relativistic hydrodynamics
simulation is beset with a number of ambiguities concerning the proper
generalization of the source density $\rho$.  Moreover, the
transformation of the second time derivative of the mass quadrupole
moment to a numerically more tractable form (time-integrated
quadrupole formula, stress formula) that does not introduce excessive
numerical noise becomes less straightforward than in the Newtonian
limit. These problems can give rise to certain pathologies in the
computed gravitational wave signals such as spurious offsets
(cp.\ Appendix~A in \citealp{dimmelmeier_02_b}), most notably (but not
exclusively) with the Newtonian stress formula.

Evolving the full space-time metric and using direct extraction would
be an obvious solution to the problem, but this would not only require
that we sacrifice the stable xCFC formulation of the metric, but
introduce other difficulties as well: Direct extraction methods can be
expected to be very vulnerable to numerical noise as the gravitational
wave signal from core-collapse supernovae is weak in the sense that it
carries away only a minuscule fraction of the total mass of the
system. Post-Newtonian generalizations of the quadrupole formula
\citep{blanchet_90} provide another possible workaround, but tend to
be considerably more cumbersome than the standard quadrupole formula.
For the extraction of gravitational waves from our core-collapse
supernova simulations, we prefer a somewhat simpler solution
that still meets the following requirements:
\begin{itemize}
\item No higher time derivatives are required for extracting the
  wave signal.
\item The gravitational wave amplitude vanishes for a stationary flow.
\item The generalization of the quadrupole formula should (and need
  only) be valid for the conditions typically encountered in
  core-collapse supernova simulations.
\end{itemize}
Due to the restriction to the supernova context, several
approximations can be made: Concerning the matter sources, we assume
all velocities to be sufficiently small to disregard retardation
effects. Concerning the space-time metric, we consider only small
perturbations on the background of a CFC metric $g_{\mu\nu}$, which is
given in terms of the lapse function $\alpha$, the shift vector
$\beta^i$, the conformal factor $\phi$, and the flat-space three
metric $\tilde{\gamma}_{ij}$ as
\begin{equation}
\label{eq:metric}
g_{\mu\nu}=
\left(
\begin{array}{cc}
 -\alpha^2 + \beta_i \beta^i  & \beta^i \\
 \beta^i & \phi^4 \tilde{\gamma}^{ij} \\
\end{array}
\right).
\end{equation}
For deriving our modified quadrupole formula, we use Cartesian
spatial coordinates so that $\tilde{\gamma}^{ij} = \delta^{ij}$, and
only convert to spherical polar coordinates to obtain the final formulae
(Equations \ref{eq:final_stress_formula} and
\ref{eq:relativistic_tqf}).
We also assume that the space-time is
almost stationary with vanishing shift,
\begin{equation}
\label{eq:metric_assumptions_1}
\frac{\pd \alpha}{\pd t} \approx 0,
\quad
\frac{\pd \phi}{\pd t} \approx 0,
\quad
\beta^i \approx 0.
\end{equation}
Furthermore, we observe empirically that for the typical field
strength reached in the proto-neutron star environment, the lapse
function and the conformal factor obey the relation
\begin{equation}
\label{eq:metric_assumptions_2}
\alpha \phi^2 \approx 1,
\end{equation}
which is a direct consequence of the fact that the field strength is
moderate and that the Eulerian energy density $E$ is the dominating
source term in both the non-linear Poisson equations for $\alpha
\phi$ and $\phi$ (cp.\ paper~I). 

With these approximations for the space-time metric, it can easily be 
verified that the harmonic gauge condition $\pd/\pd x^\nu (\sqrt{-g}
g^{\mu\nu})=0$ is satisfied. We can therefore use the relaxed field
equations \citep{straumann,pati_00} to compute gravitational
amplitudes perturbatively on the background of the CFC metric. The
relaxed field equations in geometrized units read
\begin{equation}
\label{eq:relaxed_field_equation}
\square \tilde{g}^{\mu\nu} = 16 \pi G (-g) \left(T^{\mu\nu} + \tau_\mathrm{LL}^{\mu\nu}\right)
+ \mathcal{D} (\tilde{g}^{\mu\nu}),
\end{equation}
with $g=\det (g_{\mu\nu})$, $\tilde{g}^{\mu\nu}=(-g)
g^{\mu\nu}$. Here, $T^{\mu\nu}$ and $\tau_\mathrm{LL}^{\mu\nu}$ are
the stress-energy tensor of the matter and the Landau-Lifschitz
pseudo-tensor for the gravitational field, respectively, and
$\mathcal{D}$ is a non-linear differential operator. For
  our derivation of a modified quadrupole formula, it is convenient to
  follow \citet{straumann} and to introduce
  $h^{\mu\nu}=\eta^{\mu\nu}-\sqrt{-g} g^{\mu\nu}$ (where
  $\eta^{\mu\nu}$ is the flat-space Minkowski metric).

The relaxed field equations are amenable to an iterative solution by
means of the Green's function method, i.e.\ solutions for
$h^{\alpha\beta}$ can be obtained \citep{straumann} by successively
computing
\begin{equation}
\label{eq:green_1}
h^{\mu\nu}(t,\mathbf{x})=
4 \int \frac{\tau^{\mu\nu}(t'=t-|\mathbf{x}-\mathbf{x}'| ,\mathbf{x'})}{|\mathbf{x}-\mathbf{x}'|} \ud^3 x',
\end{equation}
and updating the source $\tau^{\mu\nu}=(-g)
(T^{\mu\nu}+\tau_\mathrm{LL}^{\mu\nu})+ \mathcal{D} (\tilde{g}^{\mu\nu})/(16 \pi G)$ in the process. In general, this
procedure is beset with a number of difficulties (such as the
treatment of non-compact sources and the occurrence of divergences in
higher-order approximations), but it can readily be used to obtain an
estimate for the gravitational wave amplitude for the matter signal in
our case: As in the usual derivation of the Newtonian quadrupole
formula, we assume that the source is small compared to the wavelength
and that retardation effects can be neglected, in which case
Equation~(\ref{eq:green_1}) can be written
as\footnote{
We emphasize that unlike Equation~(\ref{eq:green_1}),
Equation~(\ref{eq:green_2}) can no longer be used to compute the gravitational
wave signal due to anisotropic neutrino emission, which must include
retardation effects as neutrinos propagate with the same velocity
as the gravitational waves themselves.
}
\begin{equation}
\label{eq:green_2}
h^{\mu\nu}(t,\mathbf{x}) \approx \frac{4}{|\mathbf{x}|}
\int \tau^{\mu\nu}(t'=t-|\mathbf{x}| ,\mathbf{x'})\, \ud^3 x'.
\end{equation}
By projecting $h^{\mu\nu}$ onto the subspace of symmetric
transverse-traceless (STT) tensors, one obtains the angle-dependent
gravitational wave amplitude, which can then be decomposed into
pure-spin tensor harmonics \citep{mathews_62,zerilli_70,thorne_80} of
degree $\ell=2$. These are already completely determined by the purely
spatial components $h^{ij}$.

At this point, the complicated metric-dependent terms in $\tau^{\mu\nu}$
are the major obstacle for formulating a simple integral expression
for the gravitational wave amplitude. However, using the fact that
$\tau^{\mu\nu}$ is divergence-free ($\pd_\nu \tau^{\mu\nu}=0$),
the integral in Equation~(\ref{eq:green_2}) can be transformed
into an integral over $\tau^{0j}$ or $\tau^{00}$ at the cost of
introducing time derivatives:
\begin{equation}
h^{ij}(t,\mathbf{x})=
\frac{4}{|\mathbf{x}|} \frac{\pd }{\pd t}
\int x^i \tau^{0j}(t-|\mathbf{x}| ,\mathbf{x'}) \, \ud^3 x'
=
\frac{4}{|\mathbf{x}|} \frac{\pd^2}{\pd t^2}
\int x^i x^j \tau^{00}(t-|\mathbf{x}| ,\mathbf{x'}) \, \ud^3 x'.
\end{equation}
The formula for $h^{ij}$ containing the time derivative of $\tau^{0j}$
is particularly useful because it does not contain purely
gravitational contributions due to our assumptions for the metric
functions, and can therefore be expressed in terms of the covariant
Eulerian three-momentum density $S_j=\rho h W^2 v_j$ (with
$h$ and $W$ being the relativistic specficic enthalpy including
rest mass contributions and the Lorenz factor, respectively) as
\begin{equation}
\tau^{0j} = (-g) T^{0j} = \alpha^2 \phi^{12} (\alpha^{-1} \phi^{-4}S_j)
= \alpha \phi^8 S_j= \phi^6 S_j= \sqrt{\gamma} S_j,
\end{equation}
which is exactly the conserved quantity in our formulation of the
equations of hydrodynamics \citep{banyuls_97}. 
The reader should note that our assumption of a diagonal space-time metric
and Cartesian spatial coordinates allows us to raise and lower indices
simply by multiplying with a scalar factor. We repeatedly and implicitly make
uses of this in the following equations.
Plugging in the
expression for $\pd \sqrt{\gamma} S_j / \pd t$ (cp.\ also
Equation~\ref{eq:gr_euler}), we now obtain
\begin{equation}
\label{eq:basic_tqf}
h^{ij}(t,\mathbf{x})
=
\frac{4}{|\mathbf{x}|} 
\int x^i
\left(\frac{\pd \sqrt{\gamma} S_j}{\pd t}\right)_\mathrm{hyd}
(t-|\mathbf{x}| ,\mathbf{x'})
\, \ud^3 x'
=
\frac{4}{|\mathbf{x}|} 
\int x^i
\left [
\frac{\sqrt{-g}}{2} T^{\mu\nu}\frac{\pd g_{\mu\nu}}{\pd x^j}
-
\frac{\pd \sqrt{-g} \left( S_j\hat{v}^k+ P \delta_j^k\right)}{\pd x^k}
\right ]
(t-|\mathbf{x}| ,\mathbf{x'}) \, \ud^3 x'.
\end{equation}
Here, $P$ is the pressure, and $\hat{v}^i=v^i-\alpha^{-1}\beta^i 
\approx v^i$ (according to our assumption of a negligible shift vector).
Note that the time derivative of $\sqrt{\gamma} S_j$ includes only the
(hydrodynamical) changes due to advection, pressure gradients and
gravitational source terms, \emph{but no neutrino momentum source
  term} $\sqrt{\gamma} \dot{S}_{j,\nu}$:
\begin{equation}
\left(\frac{\pd \sqrt{\gamma} S_j}{\pd t} \right)_\mathrm{hyd}
=
\frac{\pd \sqrt{\gamma} S_j}{\pd t}
-\sqrt{\gamma} \dot{S}_{j,\nu}.
\end{equation}
As $\sqrt{\gamma} \dot{S}_{j,\nu}$ is exactly balanced by a source
term for the neutrino energy-momentum tensor, momentum exchange
between the matter and the neutrinos does not contribute as a source
for gravitational waves.

While formula (\ref{eq:basic_tqf}) for $h^{ij}$ is already amenable to a numerical
treatment (no time derivatives are required), it can be cast into a
form corresponding more closely to the Newtonian stress formula. Using
partial integration, the term containing the flux divergence can be
eliminated again,
\begin{equation}
\label{eq:momentum_formula}
h^{ij}(\mathbf{x})=
\frac{4}{|\mathbf{x}|} 
\left[
\int
x^i\frac{\sqrt{-g}}{2} T^{\mu\nu}\frac{\pd g_{\mu\nu}}{\pd x^j}
+
\sqrt{-g} \left(S_j\hat{v}^i +P \delta_j^i\right)\, \ud^3 x'
\right],
\end{equation}
where we have suppressed the arguments $t$ and $t'$ for the sake of
simplicity.  Upon extracting the symmetric transverse-tracefree (STT) part
of the right-hand side, the pressure term drops out,
\begin{equation}
\label{eq:stress_formula_cartesian}
h_\mathrm{TT}^{ij}(\mathbf{x})=
\frac{4}{|\mathbf{x}|} 
\left(
\int
x^i\frac{\sqrt{-g}}{2} T^{\mu\nu}\frac{\pd g_{\mu\nu}}{\pd x^j}
+
\sqrt{-g} S_j\hat{v}^i \, \ud^3 x'.
\right)_\mathrm{STT}.
\end{equation}
The resemblance to the Newtonian stress formula,
\citep{nakamura_89,blanchet_90}
\begin{equation}
h_\mathrm{TT}^{ij}(\mathbf{x})=
\frac{4}{|\mathbf{x}|} 
\left(
\int
-x^i \rho \frac{\pd \Phi}{\pd x_j}
+
\rho v_j v^i \, \ud^3 x',
\right)_\mathrm{STT}
\end{equation}
is evident. The decomposition of the gravitational radiation field
given by Equation~(\ref{eq:stress_formula_cartesian}) into pure-spin
harmonics proceeds exactly as in the Newtonian case. In the case of
axisymmetry, only the ``electric'' quadrupole with $m=0$ is present,
and its amplitude $A_{20}^{\mathrm{E}2}$ can be expressed in spherical
polar coordinates as,
\begin{eqnarray}
\nonumber
A_{20}^{\mathrm{E}2}&=&
\frac{32 \pi^{3/2}}{\sqrt{15}}
\iint 
\alpha \phi^6 r^2 \sin \theta
\left[
S_r \hat{v}^r  \left(3 \cos^2 \theta - 1 \right)
+S_\theta \hat{v}^\theta \left(2 - 3 \cos^2 \theta \right)
- S_\varphi \hat{v}^\varphi
-6 r S_r \hat{v}^\theta \sin \theta \cos \theta
\right.
\\
&&
\left.
-\dot{S}_{r,\mathrm{grav}} r \left(3 \cos^2\theta-1\right)
+3 \dot{S}_{\theta,\mathrm{grav}} r \sin\theta \cos\theta
\right]
\,\ud \theta
\, \ud r,
\label{eq:final_stress_formula}
\end{eqnarray}
where $\dot{S}$ designates the gravitational source term in
the momentum equation:
\begin{equation}
\dot{S}_{i,\mathrm{grav}}=
\frac{1}{2} T^{\mu\nu}\frac{\pd g_{\mu\nu}}{\pd x^i}.
\end{equation}
Alternatively, one can retain the time-derivative $\pd \sqrt{\gamma}
S_j/\pd t$ in Equation~(\ref{eq:basic_tqf}) and formulate a modified
version of the time-integrated quadrupole formula, but the neutrino
momentum source term needs to be handled with care in this case. In
order to avoid double counting, these source terms must be subtracted
from the time derivative of $\sqrt{\gamma} S_j$, and we therefore
obtain (after converting to spherical polar coordinates again)
\begin{eqnarray}
\nonumber
A_{20}^{\mathrm{E}2}&=&
\frac{32 \pi^{3/2}}{\sqrt{15}}
\frac{\pd}{\pd t}
\iint 
\phi^6 r^3 \sin \theta
\left[S_r \left(3 \cos^2\theta -1 \right) +
3 r^{-1} S_\theta \sin\theta \cos\theta\right]
\, \ud \theta \, \ud r
\\
&&
-\frac{32 \pi^{3/2}}{\sqrt{15}}
\iint
\phi^6 r^3 \sin \theta
\left[\dot{S}_{r,\nu}, \left(3 \cos^2\theta -1 \right) 
+ 3 r^{-1} \dot{S}_{\theta,\nu} \sin\theta \cos\theta\right]
\, \ud \theta \, \ud r .
\label{eq:relativistic_tqf}
\end{eqnarray}

Since our modified stress formula has been obtained through
\emph{exact} transformations of Equation~(\ref{eq:momentum_formula}), which
contains the time derivative of $S_j$ under the integral, it does not
produce any unphysical offset in the gravitational wave signal for
stationary configurations with $\pd S_j/\pd t=0$ in the limit of
infinite spatial resolution.\footnote{For finite resolution, the identity
  $\int \tau^{ij}\, \ud V= \int x^j \tau^{0i}\, \ud V = \int
  \tau^{00} x^i x ^j\, \ud V$ does not hold exactly due to
  discretization errors, but in practice the wave signal is hardly
  affected by this problem.}  It also avoids the contamination of the
matter signal by neutrino momentum source terms since it allows for a
nice separation of the source density for the wave equation into terms
due to the hydrodynamic momentum flux and gravitational source terms.
However, the contribution from neutrino stresses is small in practice,
and may even be neglected in Equation~(\ref{eq:momentum_formula}) with little
loss of accuracy.

\section{Gravitational Wave Signal from Aspherical Shock Propagation}
\label{app:gw_shock}
As mentioned in Section~\ref{sec:general_features}, it is possible to
formulate a simple analytic expression for the contribution of
asymmetric shock propagation to the gravitational wave signal in the
limit of a relatively small deformation of the shock. To this end, we
perform the integral in the quadrupole formula
(\ref{eq:relativistic_tqf}) over a narrow region around the shock
between $r_a(\theta)$ and $r_b(\theta)$ in radius such as to neglect
slower, secular variations in the pre- and post-shock conditions. For
the sake of simplicity, we disregard relativistic kinematics and
strong-field effects, setting $\alpha=\phi=1$ and $S_i=\rho v_i$ for
the momentum density. As we confine ourselves to a weakly deformed
shock, we also neglect the non-radial component $v_\theta$ of the
post-shock velocity\footnote{Non-radial velocities occur behind a
  non-spherical shock, but a more detailed analysis of the jump
  conditions for oblique shocks reveals that this leads only to
  higher-order corrections to the resulting gravitational wave
  amplitude.} and obtain the contribution of the shock to
$A_{20,\mathrm{shock}}^\mathrm{E2}$ in terms of the (angle-dependent)
shock radius $r_\mathrm{sh}(\theta)$, and the radial pre-
and post-shock velocities ($v_\mathrm{r,p}$, $v_\mathrm{r,s}$) and
densities ($\rho_\mathrm{p}$, $\rho_\mathrm{s}$),
\begin{eqnarray}
\nonumber
  A_{20,\mathrm{shock}}^\mathrm{E2} 
  &\approx &
  \frac{32 \pi^{3/2}}{\sqrt{15}}
  \frac{\ud}{\ud t}
  \int\limits_0^\pi
  \sin \theta 
  \left[
  \int\limits_{r_a}^{r_\mathrm{sh}(\theta)} r^3 
  \rho v_r (3 \cos^2 \theta - 1)
  \, \ud r 
  +
  \int\limits_{r_\mathrm{sh}(\theta)}^{r_b} r^3 
  \rho v_r (3 \cos^2 \theta - 1)
  \, \ud r \right]
  \, \ud \theta 
\\
  &=&
  \frac{32 \pi^{3/2}}{\sqrt{15}}
  \frac{\ud}{\ud t}
  \int\limits_0^\pi
  r_\mathrm{sh}^3(\theta) 
  \sin \theta (3 \cos^2 \theta - 1)
  \left(
  \rho_\mathrm{s} v_{r,\mathrm{s}}- \rho_\mathrm{p} v_{r,\mathrm{p}} 
  \right)
  \, \ud \theta.
\end{eqnarray}
By exploiting the Rankine-Hugoniot jump conditions, the difference of
the post-shock and pre-shock mass flux can be eliminated and the shock
velocity $v_\mathrm{sh}$ appears instead:
\begin{equation}
  \rho_\mathrm{s} v_{r,\mathrm{s}}- \rho_\mathrm{p} v_{r,\mathrm{p}} 
  =
  \left(\rho_\mathrm{s} - \rho_\mathrm{p}\right) v_\mathrm{sh}
  =
  \left(\rho_\mathrm{s} - \rho_\mathrm{p}\right) \dot{r}_\mathrm{sh}.
\end{equation}
Thus, we obtain
\begin{equation}
\label{eq:gw_shock_exact}
  A_{20,\mathrm{shock}}^\mathrm{E2} \approx
  \frac{32 \pi^{3/2}}{\sqrt{15}}
  \frac{\ud}{\ud t}
  \int\limits_0^\pi \left(\rho_\mathrm{s} (\theta)-\rho_\mathrm{p} (\theta)\right)
  r_\mathrm{sh}^3(\theta) \dot{r}_\mathrm{sh}^2(\theta)
   (3 \cos^2 \theta - 1) \sin \theta
  \, \ud \theta.
\end{equation}
$r_\mathrm{sh}(\theta)$ can be decomposed into spherical harmonics,
or, in the case of axisymmetry, into Legendre polynomials,
\begin{equation}
r_\mathrm{sh}(\theta)=\sum_{\ell=0} P_\ell(\cos \theta) a_\ell.
\end{equation}
Assuming a power-law dependence of the pre-shock density $\rho \propto
r^{\gamma}$ and a fixed ratio $\beta$ of the post-shock and pre-shock
density, we can express the angle-dependence of $\rho_s (\theta)$ and
$\rho_p (\theta)$ in terms of $a_\ell$ as well:
\begin{equation}
\rho_\mathrm{s} (\theta)-\rho_\mathrm{p} (\theta)
=
(\beta-1)\bar{\rho}_\mathrm{p} \left(\frac{r_\mathrm{sh}(\theta)}{a_0}\right)^\gamma
=
(\beta-1)\bar{\rho}_\mathrm{p}
\left[
1+\gamma \sum_{\ell=1} P_\ell(\cos \theta) a_\ell a_0^{-1}
+\frac{\gamma (\gamma+1)}{2}
\left(\sum_{\ell=1} P_\ell(\cos \theta) a_\ell a_0^{-1} \right)^2
+\ldots
\right].
\end{equation}
Here, $\bar{\rho}_\mathrm{p}$ is an appropriate average
  value of $\rho_\mathrm{p}(\theta)$, and the ratio
  $(r_\mathrm{sh}(\theta)/a_0)^\gamma$ has been expanded into a power
  series. The resulting expression for
  $A_{20,\mathrm{shock}}^\mathrm{E2}$ can be expanded into a power
  series in the higher multipoles $a_1,a_2,\ldots$ of the shock
  position. Retaining only the linear terms and converting back to
  non-geometrized units, we finally obtain:
\begin{equation}
\label{eq:gw_shock}
  A_{20,\mathrm{shock}}^\mathrm{E2} \approx
  \frac{256 \pi^{3/2}}{5 \sqrt{15}}
  \bar{\rho}_\mathrm{p}
  \left(\beta -1 \right)
  a_0^3 \left[\left(4+\gamma\right) a_2 \dot{a}_0+\dot{a}_2 a_0\right].
\end{equation}

\section{Ledoux Criterion, Brunt-V\"ais\"al\"a and Plume Frequency in General Relativity}
\label{app:bv_gr}
It may not be immediately obvious to the reader how buoyancy
effects are to be treated in GR, and how, e.g., the analysis of the
deceleration of convective downdrafts at the boundary of the convection
zone presented by \citet{murphy_09} carries over to the relativistic
case. We therefore briefly recapitulate the principles governing the motion of
buoyancy-driven bubbles in the framework of GR hydrodynamics. Although
the problem of convective stability has been studied in general
relativity \citep{thorne_66}, we find it advisable to explicitly
derive the equations for the particular gauge and frame-of-reference
used in our paper in order to eliminate a possible source of
confusion.

We start from the relativistic momentum equation in the $3+1$ formulation
\citep{banyuls_97},
\begin{equation}
\label{eq:gr_euler}
\frac{\pd \sqrt{\gamma} \rho h W^2 v_i}{\pd t}
+
\frac{\pd \sqrt{-g} \left[\rho h W^2 v_i \left(v^j - \beta^j/\alpha\right)+ \delta_i^j P\right]}{\pd x^j}
=
\frac{\sqrt{-g}}{2}T^{\mu\nu}\frac{\pd g_{\mu\nu}}{\pd x^j}.
\end{equation}
Here, $\rho$, $h$, $W$, $P$, $v_i$, $T^{\mu\nu}$ denote the rest-mass
density, the specific enthalpy, the Lorentz factor, the pressure, the
three-velocity and the stress-energy-tensor of the fluid,
respectively. The four-metric $g_{\mu\nu}$ is given in terms
of the lapse function $\alpha$, the shift vector $\beta^i$,
the conformal factor $\phi$, and the flat-space three metric
$\tilde{\gamma}_{ij}$ by Equation~(\ref{eq:metric}), which is repeated
here for convenience:
\begin{equation}
g_{\mu\nu}=
\left(
\begin{array}{cc}
 -\alpha^2 + \beta_i \beta^i  & \beta^i \\
 \beta^i & \phi^4 \tilde{\gamma}^{ij} \\
\end{array}
\right).
\end{equation}
Finally, $\gamma$ and $g$ denote the determinants of the three-metric $\gamma_{ij}$ and the
four-metric $g_{\mu\nu}$.

We now consider the equation of motion for a convective blob of matter
in a hydrostatic background medium which is displaced by a short
distance $\delta r$ from its original position and maintains
pressure equilibrium with its surroundings. The acceleration of
the blob due to pressure gradients is given by the second term on the
LHS of Equation~(\ref{eq:gr_euler}), which can be replaced by the
gravitational source term \emph{for the background medium} using the
condition of hydrostatic equilibrium. The equation of motion
(including the gravitational source term) is therefore given by
\begin{equation}
\left(\frac{\pd \sqrt{\gamma} \rho h W^2 v_i}{\pd t}\right)_\mathrm{blob}
=
\frac{\alpha \sqrt{\gamma}}{2}(T_\mathrm{blob}^{\mu\nu}-T_\mathrm{bg}^{\mu\nu})\frac{\pd g_{\mu\nu}}{\pd x^j},
\end{equation}
where the subscripts denote whether $T^{\mu\nu}$ is evaluated for the
blob or the background medium (``bg''). In order to simplify this
result, we assume slow subsonic motion ($v \ll c_s$), and retain
density and energy perturbations only in the buoyancy term on the RHS
(the Boussinesq approximation). The only remaining contribution on the
RHS then contains $T_\mathrm{blob}^{00}-T_\mathrm{bg}^{00}$, and we obtain
\begin{equation}
\rho h \frac{\ud v_\mathrm{r,blob} \phi^2}{\ud t}
=- \delta (\rho +\rho \epsilon) \frac{\pd \alpha}{\pd r},
\end{equation}
where $\delta (\rho +\rho \epsilon)$ denotes the (energy-)density
contrast between the blob and the background medium (with $\epsilon$
denoting the specific internal energy). For a small radial
displacement $\delta r$ from the initial position, $\delta (\rho +\rho
\epsilon)$ can be expressed as $C_L \delta r$, $C_L$ being the
(relativistic) Ledoux criterion (cp.\ the ``Schwarzschild discriminant''
of \citealt{thorne_66}),
\begin{equation}
C_\mathrm{L}=\frac{\pd \rho (1+\epsilon)}{\pd r}-\left(\frac{\ud
  \rho (1+\epsilon)}{\ud P}\right)_{s,Y_e=\mathrm{const}.}\frac{\pd P}{\pd r}=
\frac{\pd \rho (1+\epsilon)}{\pd r}-\frac{1}{c_s^2}\frac{\pd P}{\pd r}.
\end{equation}
The equation of motion for the blob can thus be written as
\begin{equation}
\frac{\ud v_\mathrm{r,blob}}{\ud t} =- \frac{\phi^{-2}
  C_\mathrm{L}}{\rho h} \frac{\pd \alpha}{\pd r} \delta r,
\end{equation}
or
\begin{equation}
\label{eq:ode_blob}
\ddot{r}
= \alpha \phi^{-2} \frac{\ud v_\mathrm{r,blob}}{\ud t}
= - \frac{\alpha C_\mathrm{L}}{\rho h \phi^4} \frac{\pd \alpha}{\pd r} \delta r.
\end{equation}
The blob therefore oscillates with an angular frequency $N$ given by
\begin{equation}
N^2 = \frac{\alpha C_\mathrm{L}}{\rho h \phi^4} \frac{\pd \alpha}{\pd
  r}.
\end{equation}
This is the relativistic Brunt-V\"ais\"al\"a-frequency. It should be
noted that we have measured $N$ in coordinate time and not in the
frame of an observer comoving with the fluid. As the spacetime is
stationary to very good approximation in our case, this implies that
$N$ is also the angular frequency seen by an observer at
infinity.\footnote{This can easily be seen by considering geodesics
  starting from the position of the blob at two different coordinate
  times $t$ and $t+\delta t$. Because of the stationarity of the
  metric, the geodesics can be transformed into each other by applying
  a time translation $\delta t$ \emph{everywhere}. An observer at
  infinity therefore measures the same time difference $\delta t$.}

For determining the ``plume frequency'' and the penetration depth
$D_p$ used in the model of \citet{murphy_09}, we estimate the
turnaround time $T$ and $D_p$ for a convective blob overshooting into
a stable layer assuming a constant value of $N$ equal to that at the
turning point during the decelerating process.  Moreover, we assume
that the initial density contrast to the background medium at the
boundary of the convective and non-convective regions is negligible
compared to the density contrast after penetration into the
non-convective region.  Otherwise an additional acceleration term
would have to be included in the equation of motion.
Equation~(\ref{eq:ode_blob}) can then be solved using $\delta r(0)=0$
and $\dot{r}(0)=-\alpha \phi^2 v_r$ ($v_r$ being the plume velocity in
the convective region) as initial conditions,
\begin{equation}
\delta r(t)=\alpha \phi^2 v_r \sin N t.
\end{equation}
The penetration depth is therefore given by
\begin{equation}
D_p=\alpha \phi^2 v_r,
\end{equation}
and the characteristic (ordinary) frequency $f_p$ for the motion of
the decelerating blob corresponding to the angular frequency is just
\begin{equation}
f_p=\frac{N}{2 \pi}.
\end{equation}

\end{document}